%% file: main.tex
\pdfoutput=1

\documentclass[12pt,a4paper]{article}

\usepackage{ifthen} 
\newboolean{pdflatex}
\setboolean{pdflatex}{true} 

\newboolean{articletitles}
\setboolean{articletitles}{true} 

\newboolean{uprightparticles}
\setboolean{uprightparticles}{false} 

\def\paperauthors{LHCb collaboration} 
\def\paperasciititle{Measurement of the branching fraction of the Bd->DsPi decay} 
\def\papertitle{Measurement of the branching fraction of the $\BdDsPi$ decay} 
\def\paperkeywords{{High Energy Physics}, {LHCb}} 
\def\papercopyright{\the\year\ CERN for the benefit of the LHCb collaboration} 
\def\paperlicence{CC BY 4.0 licence}
\def\paperlicenceurl{https://creativecommons.org/licenses/by/4.0/}

\newcommand{\BdDsPi}{\decay{\Bd}{\Dsp\pim}}
\newcommand{\BdDPi}{\decay{\Bd}{\Dm\pip}}
\newcommand{\BsDsPi}{\decay{\Bsb}{\Dsp\pim}}
\newcommand{\BsDsK}{\decay{\BsorBsbar}{\Dsp\Km}}
\newcommand{\BdDK}{\decay{\Bd}{\Dm\Kp}}
\newcommand{\LbLcPi}{\decay{\Lb}{\Lc\pim}}
\newcommand{\BdDstPi}{\decay{\Bd}{\Dstarm\pip}}
\newcommand{\BdDRho}{\decay{\Bd}{\Dm\rhop}}
\newcommand{\BsDsstPi}{\decay{\Bsb}{\Dssp\pim}}
\newcommand{\BsDsRho}{\decay{\Bsb}{\Dsp\rhom}}
\newcommand{\DKPiPi}{\decay{\Dm}{\Kp\pim\pim}}
\newcommand{\DsKKPi}{\decay{\Dsp}{\Kp\Km\pip}}

\newcommand{\Vub}{V_{ub}}
\newcommand{\aNF}{a_{\text{NF}}}

\usepackage{float}
\usepackage{subcaption}

\makeatletter
\g@addto@macro\bfseries{\boldmath}
\makeatother

\input{preamble}
\input{resultFile}
\usepackage{longtable} 
\usepackage{booktabs}
\usepackage{siunitx}
\usepackage[bottom]{footmisc}

\begin{document}

\renewcommand{\thefootnote}{\fnsymbol{footnote}}
\setcounter{footnote}{1}

\input{title-LHCb-PAPER}


\renewcommand{\thefootnote}{\arabic{footnote}}
\setcounter{footnote}{0}

\cleardoublepage


\pagestyle{plain} 
\setcounter{page}{1}
\pagenumbering{arabic}


%

\input{introduction}

\input{detector}

\input{selection}

\input{parametrisations}

\input{massFits}

\input{systematics}

\input{results}

\input{summary}


\input{acknowledgements}

\addcontentsline{toc}{section}{References}
\bibliographystyle{LHCb}
\bibliography{main,standard,LHCb-PAPER,LHCb-CONF,LHCb-DP,LHCb-TDR}

\newpage
\input{LHCb_Authorship_2020-Oct-22}




\end{document}

%% file: preamble.tex
\usepackage[top=1in, bottom=1.25in, left=1in, right=1in]{geometry}

\usepackage{microtype}
\usepackage{lineno}  
\usepackage{xspace} 
\usepackage{caption} 

\usepackage{graphicx}  
\usepackage{color}
\usepackage{colortbl}
\graphicspath{{./figs/}} 

\usepackage{amsmath} 
\usepackage{amssymb}
\usepackage{amsfonts}
\usepackage{upgreek} 

\newcommand*\patchAmsMathEnvironmentForLineno[1]{%
\expandafter\let\csname old#1\expandafter\endcsname\csname #1\endcsname
\expandafter\let\csname oldend#1\expandafter\endcsname\csname
end#1\endcsname
 \renewenvironment{#1}%
   {\linenomath\csname old#1\endcsname}%
   {\csname oldend#1\endcsname\endlinenomath}%
}
\newcommand*\patchBothAmsMathEnvironmentsForLineno[1]{%
  \patchAmsMathEnvironmentForLineno{#1}%
  \patchAmsMathEnvironmentForLineno{#1*}%
}
\AtBeginDocument{%
\patchBothAmsMathEnvironmentsForLineno{equation}%
\patchBothAmsMathEnvironmentsForLineno{align}%
\patchBothAmsMathEnvironmentsForLineno{flalign}%
\patchBothAmsMathEnvironmentsForLineno{alignat}%
\patchBothAmsMathEnvironmentsForLineno{gather}%
\patchBothAmsMathEnvironmentsForLineno{multline}%
\patchBothAmsMathEnvironmentsForLineno{eqnarray}%
}

\usepackage{hyperxmp}

\usepackage[pdftex,
            pdfauthor={\paperauthors},
            pdftitle={\paperasciititle},
            pdfkeywords={\paperkeywords},
            pdfcopyright={Copyright (C) \papercopyright},
            pdflicenseurl={\paperlicenceurl}]{hyperref}

\usepackage[colorinlistoftodos,textsize=scriptsize]{todonotes}

\usepackage[all]{hypcap} 

\input{lhcb-symbols-def} 

\usepackage{cite} 
\usepackage{mciteplus}

%% file: lhcb-symbols-def.tex
\usepackage{xspace} 
\usepackage{upgreek}

\def\lhcb   {\mbox{LHCb}\xspace}

\def\MagUp {\mbox{\em Mag\kern -0.05em Up}\xspace}

\ifthenelse{\boolean{uprightparticles}}%
{

 \def\Ppi         {\ensuremath{\uppi}\xspace}                 
                  
 \def\Prho        {\ensuremath{\uprho}\xspace}

 \def\PDelta      {\ensuremath{\Delta}\xspace}                 
 \def\PXi         {\ensuremath{\Xi}\xspace}                 
 \def\PLambda     {\ensuremath{\Lambda}\xspace}                 
 \def\PSigma      {\ensuremath{\Sigma}\xspace}                 
 \def\POmega      {\ensuremath{\Omega}\xspace}                 
 \def\PUpsilon    {\ensuremath{\Upsilon}\xspace}

 \def\PB      {\ensuremath{\mathrm{B}}\xspace}                 
                  
 \def\PD      {\ensuremath{\mathrm{D}}\xspace}

 \def\PK      {\ensuremath{\mathrm{K}}\xspace}

 \def\Pb      {\ensuremath{\mathrm{b}}\xspace}                 
 \def\Pc      {\ensuremath{\mathrm{c}}\xspace}

 \def\Pi      {\ensuremath{\mathrm{i}}\xspace}

 \def\Pp      {\ensuremath{\mathrm{p}}\xspace}

 \def\Ps      {\ensuremath{\mathrm{s}}\xspace}                 
                  
 \def\Pu      {\ensuremath{\mathrm{u}}\xspace}

 \def\thebaroffset{0.0em}
}
{

 \def\Ppi         {\ensuremath{\pi}\xspace}                 
                  
 \def\Prho        {\ensuremath{\rho}\xspace}

 \mathchardef\PDelta="7101
 \mathchardef\PXi="7104
 \mathchardef\PLambda="7103
 \mathchardef\PSigma="7106
 \mathchardef\POmega="710A
 \mathchardef\PUpsilon="7107
                  
 \def\PB      {\ensuremath{B}\xspace}                 
                  
 \def\PD      {\ensuremath{D}\xspace}

 \def\PK      {\ensuremath{K}\xspace}

 \def\Pb      {\ensuremath{b}\xspace}                 
 \def\Pc      {\ensuremath{c}\xspace}

 \def\Pi      {\ensuremath{i}\xspace}

 \def\Pp      {\ensuremath{p}\xspace}

 \def\Ps      {\ensuremath{s}\xspace}                 
                  
 \def\Pu      {\ensuremath{u}\xspace}

 \def\thebaroffset{0.18em}
}
\newcommand{\offsetoverline}[2][\thebaroffset]{\kern #1\overline{\kern -#1 #2}}%

\makeatletter
\ifcase \@ptsize \relax
  \newcommand{\miniscule}{\@setfontsize\miniscule{4}{5}}
\or
  \newcommand{\miniscule}{\@setfontsize\miniscule{5}{6}}
\or
  \newcommand{\miniscule}{\@setfontsize\miniscule{5}{6}}
\fi
\makeatother

\DeclareRobustCommand{\optbar}[1]{\shortstack{{\miniscule (\rule[.5ex]{1.25em}{.18mm})}
  \\ [-.7ex] $#1$}}

\def\uquark    {{\ensuremath{\Pu}}\xspace}

\def\squark    {{\ensuremath{\Ps}}\xspace}

\def\cquark    {{\ensuremath{\Pc}}\xspace}

\def\bquark    {{\ensuremath{\Pb}}\xspace}

\def\pion   {{\ensuremath{\Ppi}}\xspace}
\def\piz    {{\ensuremath{\pion^0}}\xspace}
\def\pip    {{\ensuremath{\pion^+}}\xspace}
\def\pim    {{\ensuremath{\pion^-}}\xspace}
\def\pipm   {{\ensuremath{\pion^\pm}}\xspace}

\def\rhomeson {{\ensuremath{\Prho}}\xspace}

\def\rhop     {{\ensuremath{\rhomeson^+}}\xspace}
\def\rhom     {{\ensuremath{\rhomeson^-}}\xspace}

\def\kaon    {{\ensuremath{\PK}}\xspace}

\def\KorKbar {\kern \thebaroffset\optbar{\kern -\thebaroffset \PK}{}\xspace}

\def\Kp      {{\ensuremath{\kaon^+}}\xspace}
\def\Km      {{\ensuremath{\kaon^-}}\xspace}

\def\D       {{\ensuremath{\PD}}\xspace}

\def\DorDbar {\kern \thebaroffset\optbar{\kern -\thebaroffset \PD}\xspace}

\def\Dp      {{\ensuremath{\D^+}}\xspace}
\def\Dm      {{\ensuremath{\D^-}}\xspace}

\def\Dmp     {{\ensuremath{\D^\mp}}\xspace}
\def\DpDm    {\ensuremath{\Dp {\kern -0.16em \Dm}}\xspace}

\def\Dstarp  {{\ensuremath{\D^{*+}}}\xspace}
\def\Dstarm  {{\ensuremath{\D^{*-}}}\xspace}

\def\Dsp     {{\ensuremath{\D^+_\squark}}\xspace}
\def\Dsm     {{\ensuremath{\D^-_\squark}}\xspace}

\def\Dssp    {{\ensuremath{\D^{*+}_\squark}}\xspace}

\def\B       {{\ensuremath{\PB}}\xspace}
\def\Bbar    {{\ensuremath{\offsetoverline{\PB}}}\xspace}

\def\BorBbar {\kern \thebaroffset\optbar{\kern -\thebaroffset \PB}\xspace}

\def\Bd      {{\ensuremath{\B^0}}\xspace}
\def\Bdb     {{\ensuremath{\Bbar{}^0}}\xspace}
\def\BdorBdbar {\kern \thebaroffset\optbar{\kern -\thebaroffset \Bd}\xspace}

\def\Bs      {{\ensuremath{\B^0_\squark}}\xspace}
\def\Bsb     {{\ensuremath{\Bbar{}^0_\squark}}\xspace}
\def\BsorBsbar {\kern \thebaroffset\optbar{\kern -\thebaroffset \Bs}\xspace}

\def\Y#1S{\ensuremath{\PUpsilon{(#1S)}}\xspace}

\def\proton      {{\ensuremath{\Pp}}\xspace}
\def\antiproton  {{\ensuremath{\overline \proton}}\xspace}

\def\Lz          {{\ensuremath{\PLambda}}\xspace}
\def\Lbar        {{\ensuremath{\offsetoverline{\PLambda}}}\xspace}
\def\LorLbar     {\kern \thebaroffset\optbar{\kern -\thebaroffset \PLambda}\xspace}

\def\Lc          {{\ensuremath{\Lz^+_\cquark}}\xspace}
\def\Lcbar       {{\ensuremath{\Lbar{}^-_\cquark}}\xspace}

\def\Lb           {{\ensuremath{\Lz^0_\bquark}}\xspace}
\def\Lbbar        {{\ensuremath{\Lbar{}^0_\bquark}}\xspace}

\def\BF         {{\ensuremath{\mathcal{B}}}\xspace}
\def\BR         {\BF}

\newcommand{\decay}[2]{\ensuremath{#1\!\to #2}\xspace} 

\def\to                 {\ensuremath{\rightarrow}\xspace}

\def\grpsuthree {{\ensuremath{\mathrm{SU}(3)}}\xspace}

\def\CP                {{\ensuremath{C\!P}}\xspace}

\def\Vub  {{\ensuremath{V_{\uquark\bquark}}}\xspace}

\def\AT#1     {\ensuremath{A_{\mathrm{T}}^{#1}}\xspace}           

\def\C#1      {\ensuremath{\mathcal{C}_{#1}}\xspace}                       
\def\Cp#1     {\ensuremath{\mathcal{C}_{#1}^{'}}\xspace}                    
\def\Ceff#1   {\ensuremath{\mathcal{C}_{#1}^{\mathrm{(eff)}}}\xspace}        
\def\Cpeff#1  {\ensuremath{\mathcal{C}_{#1}^{'\mathrm{(eff)}}}\xspace}       
\def\Ope#1    {\ensuremath{\mathcal{O}_{#1}}\xspace}                       
\def\Opep#1   {\ensuremath{\mathcal{O}_{#1}^{'}}\xspace}                    

\newcommand{\nospaceunit}[1]{\ensuremath{\text{#1}}}       
\newcommand{\aunit}[1]{\ensuremath{\text{\,#1}}}       

\newcommand{\tev}{\aunit{Te\kern -0.1em V}\xspace}
\newcommand{\gev}{\aunit{Ge\kern -0.1em V}\xspace}
\newcommand{\mev}{\aunit{Me\kern -0.1em V}\xspace}
\newcommand{\kev}{\aunit{ke\kern -0.1em V}\xspace}
\newcommand{\ev}{\aunit{e\kern -0.1em V}\xspace}
\newcommand{\mevc}{\ensuremath{\aunit{Me\kern -0.1em V\!/}c}\xspace}
\newcommand{\gevc}{\ensuremath{\aunit{Ge\kern -0.1em V\!/}c}\xspace}
\newcommand{\mevcc}{\ensuremath{\aunit{Me\kern -0.1em V\!/}c^2}\xspace}
\newcommand{\gevcc}{\ensuremath{\aunit{Ge\kern -0.1em V\!/}c^2}\xspace}

\def\mum  {\ensuremath{\,\upmu\nospaceunit{m}}\xspace}

\def\fb   {\ensuremath{\aunit{fb}}\xspace}
\def\invfb   {\ensuremath{\fb^{-1}}\xspace}

\newcommand{\chisqip}{\ensuremath{\chi^2_{\text{IP}}}\xspace}

\def\gsim{{~\raise.15em\hbox{$>$}\kern-.85em
          \lower.35em\hbox{$\sim$}~}\xspace}
\def\lsim{{~\raise.15em\hbox{$<$}\kern-.85em
          \lower.35em\hbox{$\sim$}~}\xspace}

\def\pt         {\ensuremath{p_{\mathrm{T}}}\xspace}

\def\ptot       {\ensuremath{p}\xspace}

\def\evtgen     {\mbox{\textsc{EvtGen}}\xspace}

\def\geant      {\mbox{\textsc{Geant4}}\xspace}

\def\photos     {\mbox{\textsc{Photos}}\xspace}

\def\pythia     {\mbox{\textsc{Pythia}}\xspace}

\def\tell1  {TELL1\xspace}
\def\ukl1   {UKL1\xspace}

\newcommand{\eg}{\mbox{\itshape e.g.}\xspace}
\newcommand{\ie}{\mbox{\itshape i.e.}\xspace}


%% file: resultFile.tex
\newcommand{\BFBdDsPiPaper}{ (19.4 \pm 1.8\pm 1.3 \pm 1.2)\times 10^{-6} }

\newcommand{\RatioBFBdDsPiBFBdDPiPaper}{ (7.7 \pm 0.7 \pm 0.5 \pm 0.3)\times 10^{-3} }

\newcommand{\VubaNFExpanded}{(3.14 \pm 0.20 \pm 0.25)\times 10^{-3}}

\newcommand{\rDPiExpanded}{0.0163\pm 0.0007 \pm 0.0007 \pm 0.0033}

\newcommand{\fsfdThirteenSevenPaper}{1.020 \pm 0.013 \pm 0.021}

\newcommand{\fsfdThirteenEightPaper}{1.035 \pm 0.011 \pm 0.021}

\newcommand{\fsfdEightSevenPaper}{0.986 \pm 0.013 \pm 0.021}

\newcommand{\RSevenPaper}{0.1631 \pm 0.0018 \pm 0.0025 \pm 0.0014}

\newcommand{\REightPaper}{0.1609 \pm 0.0013 \pm 0.0024 \pm 0.0014}

\newcommand{\RThirteenPaper}{0.1665 \pm 0.0011 \pm 0.0023 \pm 0.0012}

\newcommand{\YieldBdDPiRunOnePaper}{(4.971\pm0.013)\times 10^5}

\newcommand{\YieldBdDPiRunTwoPaper}{(6.294\pm 0.016)\times 10^5}

\newcommand{\YieldBsDsPiRunOnePaper}{(3.370 \pm 0.023)\times 10^4}

\newcommand{\YieldBsDsPiRunTwoPaper}{(4.647 \pm 0.027)\times 10^4}

\newcommand{\YieldBdDsPiRunOnePaper}{(8.9 \pm 0.8)\times 10^2}

\newcommand{\YieldBdDsPiRunTwoPaper}{(1.12 \pm 0.11)\times 10^3}

\newcommand{\EffBdDPiRunOne}{0.3485\pm0.0016}
\newcommand{\EffBdDPiRunTwo}{0.4536\pm0.0016}

\newcommand{\EffBdDsPiRunOne}{0.1412\pm0.0010}
\newcommand{\EffBdDsPiRunTwo}{0.1922\pm0.0012}

%% file: title-LHCb-PAPER.tex

\begin{titlepage}
\pagenumbering{roman}

\vspace*{-1.5cm}
\centerline{\large EUROPEAN ORGANIZATION FOR NUCLEAR RESEARCH (CERN)}
\vspace*{1.5cm}
\noindent
\begin{tabular*}{\linewidth}{lc@{\extracolsep{\fill}}r@{\extracolsep{0pt}}}
\ifthenelse{\boolean{pdflatex}}
{\vspace*{-1.5cm}\mbox{\!\!\!\includegraphics[width=.14\textwidth]{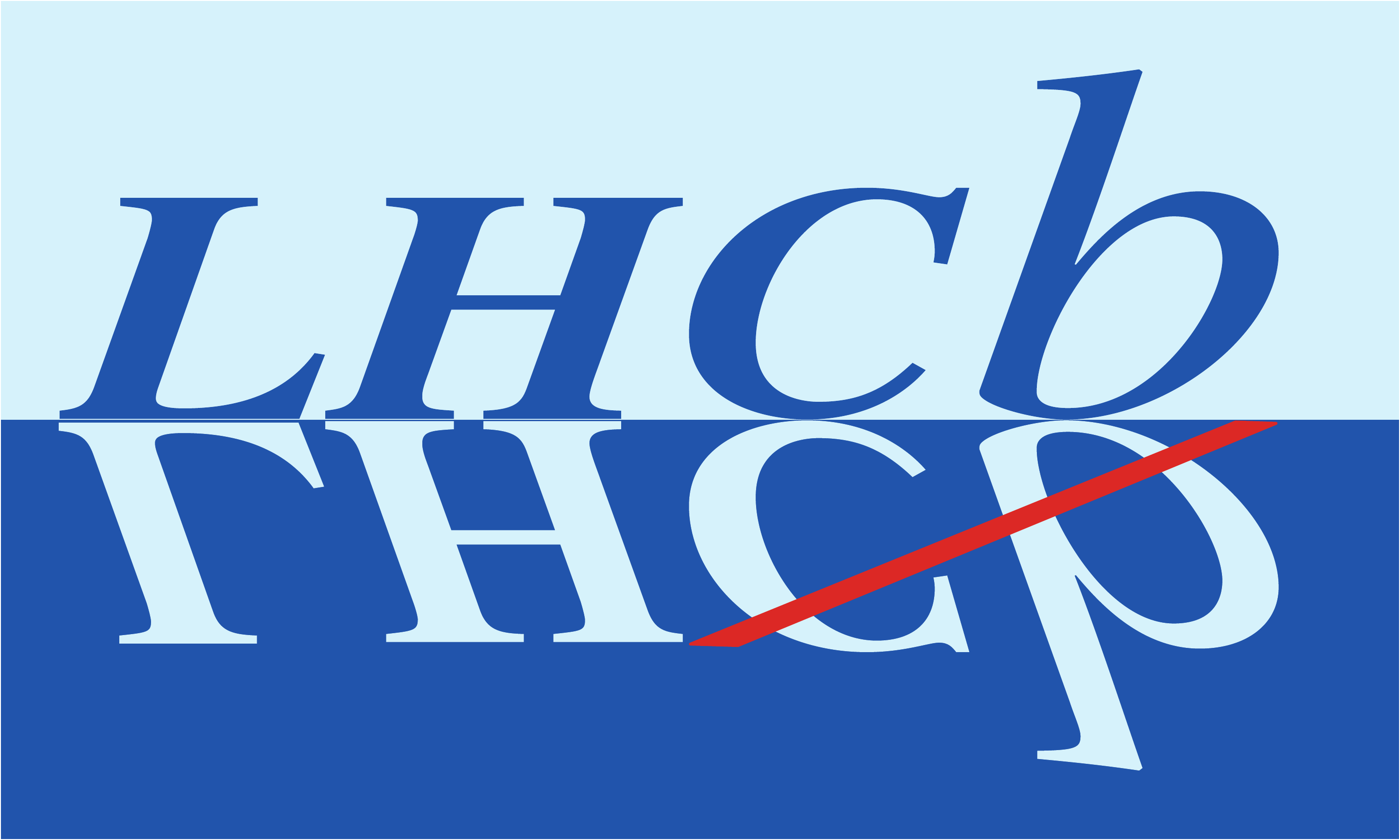}} & &}%
{\vspace*{-1.2cm}\mbox{\!\!\!\includegraphics[width=.12\textwidth]{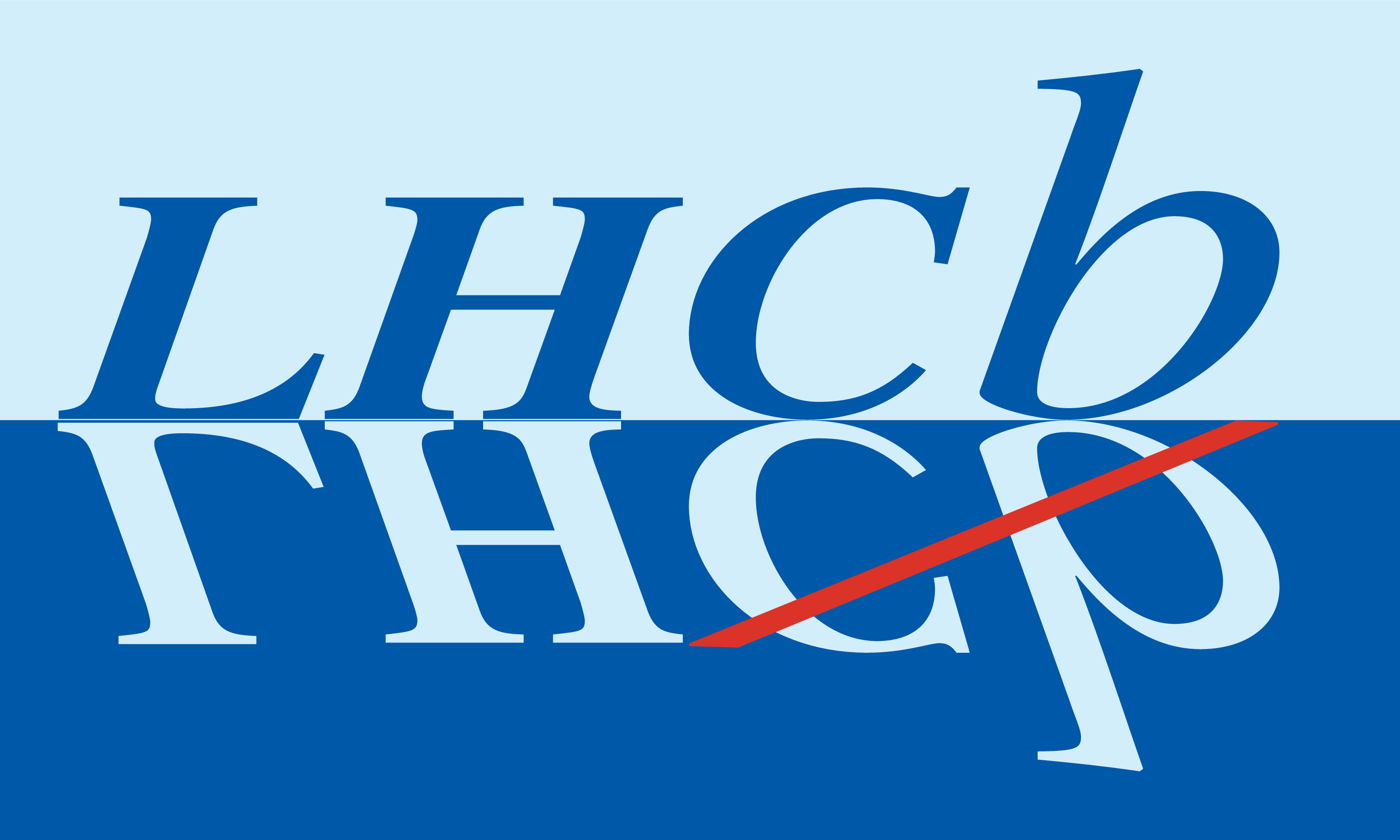}} & &}%
\\
 & & CERN-EP-2020-183 \\  
 & & LHCb-PAPER-2020-021 \\  
 & & April 19, 2021 \\ 
 & & \\
\end{tabular*}

\vspace*{3.0cm}

{\normalfont\bfseries\boldmath\huge
\begin{center}
  \papertitle 
\end{center}
}

\vspace*{1.0cm}

\begin{center}
\paperauthors\footnote{Authors are listed at the end of this paper.}
\end{center}

\vspace{\fill}

\begin{abstract}
  \noindent
    A branching fraction measurement of the $\BdDsPi$ decay is presented using proton-proton collision data collected with the \lhcb experiment, corresponding to an integrated luminosity of $5.0\invfb$. The branching fraction is found to be ${\BF(\BdDsPi) = \BFBdDsPiPaper}$, where the first uncertainty is statistical, the second systematic and the third is due to the uncertainty on the $\Bd\to\Dm\pip$, $\DsKKPi$ and $\DKPiPi$ branching fractions. This is the most precise single measurement of this quantity to date. As this decay proceeds through a single amplitude involving a $b \to u$ charged-current transition, the result provides information on non-factorisable strong interaction effects and the magnitude of the Cabibbo-Kobayashi-Maskawa matrix element $V_{ub}$. Additionally, the collision energy dependence of the hadronisation-fraction ratio $f_s/f_d$ is measured through $\BsDsPi$ and $\BdDPi$ decays.
\end{abstract}
\vspace*{2.0cm}

\begin{center}
  Published in Eur.~Phys.~J.~C81~(2021)~314
\end{center}

\vspace{\fill}

{\footnotesize 
\centerline{\copyright~\papercopyright. \href{\paperlicenceurl}{\paperlicence}.}}
\vspace*{2mm}

\end{titlepage}


\newpage
\setcounter{page}{2}
\mbox{~}
%
%
%
%

%% file: introduction.tex
\section{Introduction}
\label{sec:Introduction}
\begin{figure}[b]
\centering
\vspace*{0cm}
\includegraphics[scale=0.25]{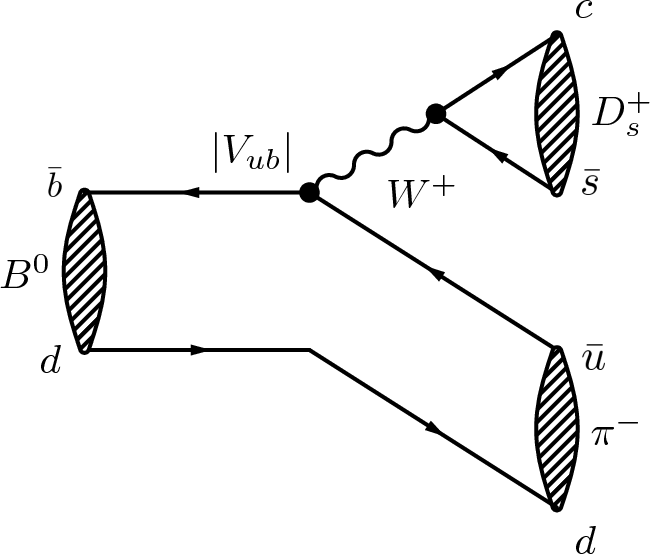}
\caption{Tree diagram of the $\BdDsPi$ decay, in which a \Bd meson decays through the weak interaction to a \Dsp meson and a charged pion. This diagram represents the only (leading order) process contributing to this decay. Strong interaction between the \Dsp meson and the pion lead to a non-factorisable contribution to the decay amplitude. }
\vspace{-0.1cm}
\label{fig:Bd2DsPi_fey}
\end{figure}

To test the Cabibbo-Kobayashi-Maskawa (CKM) sector of the Standard Model (SM), it is crucial to perform accurate measurements of the quark-mixing matrix elements. Any discrepancy among the numerous measurements of \mbox{CKM matrix} elements could reveal effects from new particles or forces beyond the SM. The knowledge of the magnitude of the matrix element $\Vub$ governing the strength of $\decay{\bquark}{\uquark}$ transitions is key in the consistency checks of the SM and its naturally motivated extensions~\cite{CKMfitter2015, UTfit-UT}. \\
\indent The hadronic $\BdDsPi$ decay\footnote{Inclusion of charge-conjugate modes is implied unless explicitly stated.} proceeds in the SM through the \decay{\bquark}{\uquark} transition as shown in Fig.~\ref{fig:Bd2DsPi_fey}. Its branching fraction is proportional to $|\Vub|^{2}$,
\begin{equation}
    \BF(\BdDsPi) =  \Phi |\Vub|^{2} |V_{cs}|^{2}|F(\Bd \rightarrow \pim)|^{2} f_{\Dsp}^{2} |\aNF|^{2},
\label{eq:factorisation}
\end{equation}
where $\Phi$ is a phase-space factor, $F(\Bd\rightarrow \pim)$ is a form factor, $f_{\Dsp}$ is the \Dsp decay constant, $V_{cs}$ is the CKM matrix element representing $\cquark \to \squark$ transitions, and $|\aNF|$ encapsulates non-factorisable effects. The form factor and the decay constant can be obtained from light-cone sum rules~\cite{formfactor,Ball:2006jz} and lattice QCD calculations~\cite{decayconstant,decayconstant2}, and since $|V_{cs}|$ is known to be close to unity, the $\BdDsPi$ branching fraction can be used to probe the product $|\Vub||\aNF|$.
The assumption of factorisation is expected to hold, \ie $|\aNF|$ is close to unity, for $B$ meson decays into a heavy and a light meson, where the $W$ emission of the decay corresponds to the light meson and the spectator quark forms part of the heavy meson. This is not the case for the $\BdDsPi$ decay, as shown in Fig~\ref{fig:Bd2DsPi_fey}, and consequently $|\aNF|$ may be significantly different from unity~\cite{Beneke:2000ry}.

The measurement of the $\BdDsPi$ branching fraction can also be used to estimate the ratio of the amplitudes of the Cabibbo-suppressed $\decay{\Bd}{\Dp\pim}$ and the Cabibbo-favoured $\decay{\Bd}{\Dm\pip}$ decays,
\begin{equation}
    r_{D\pi} = \left| \frac{A(\decay{\Bd}{\Dp\pim})}{A(\decay{\Bd}{\Dm\pip})}\right|,
\label{eq:rDPi_amplitudes}
\end{equation}
which is necessary for the measurement of charge-parity (\CP) asymmetries in \mbox{\decay{\Bd}{\Dmp\pipm}} decays~\cite{Aubert:2005yf,Aubert:2006tw,Ronga:2006hv,Bahinipati:2011yq,DeBruyn:2012jp,LHCb-PAPER-2018-009}. 
Assuming \grpsuthree flavour symmetry, Eq.~\eqref{eq:rDPi_amplitudes} can be written as~\cite{Aubert:2008zi,Das:2010be}
\begin{equation}
    r_{D\pi} = \tan{\theta_{c}}\frac{f_{\Dp}}{f_{\Dsp}}\sqrt{\frac{\BF(\BdDsPi)}{\BF(\BdDPi)}},
\label{eq:rDPi}
\end{equation}
where $\theta_{c}$ is the Cabibbo angle and $f_{\Dp}$ is the decay constant of the $\Dp$ meson. \grpsuthree symmetry breaking is caused by different non-factorisable effects in in $\BdDsPi$ and $\decay{\Bd}{\Dp\pim}$ decays.\\
\indent This article presents measurements of $\BF(\BdDsPi)$ and $r_{D\pi}$  using proton-proton ($pp$) collision data collected with the \lhcb detector at centre-of-mass energies of 7, 8 and 13 \tev corresponding to an integrated luminosity of $5 \invfb$. The data samples recorded in the years 2011 and 2012 (2015 and 2016) at 7 and 8 (13) \tev will be referred to as \mbox{Run 1 (Run 2)}. The $\BdDsPi$ branching ratio is measured relative to the $\BdDPi$ normalisation channel, which is well measured and experimentally similar to the $\BdDsPi$ decay. The $\BdDsPi$ ($\BdDPi$) candidates are reconstructed via the \decay{\Dsp}{\Kp\Km\pip} (\decay{\Dm}{\Kp\pim\pim}) decay. The branching fraction of the $\BdDsPi$ decay is determined by
\begin{equation}
    \BF(\BdDsPi) = \BF(\BdDPi) \dfrac{N_{\BdDsPi}}{N_{\BdDPi}}\dfrac{\epsilon_{\BdDPi}}{\epsilon_{\BdDsPi}}\dfrac{\BF(\DKPiPi)}{\BF(\DsKKPi)},
\label{eq:BF_Bd2DsPi}
\end{equation}
where $N_{\rm{X}}$ denotes the selected candidate yield and $\epsilon_{\rm{X}}$ the related efficiency for the decay mode X. In this measurement, extended maximum-likelihood fits to unbinned invariant mass distributions are performed in order to obtain the yields, while the efficiencies are obtained from simulated events and using calibration data samples. \\
\indent The relative production of \Bs and \Bd mesons, described by the ratio $f_s/f_d$ where $f_s$ and $f_d$ are the $\Bs$ and $\Bd$ hadronisation fractions, is shown to slightly depend on the $pp$ collision energy~\cite{LHCb-PAPER-2019-020}. The efficiency-corrected yield ratio $\cal R$, 
\begin{equation}
    \mathcal{R} \equiv \dfrac{N_{\BsDsPi}}{N_{\BdDPi}}\dfrac{\epsilon_{\BdDPi}}{\epsilon_{\BsDsPi}} \propto \dfrac{f_s}{f_d}, \label{eq:fsfd_prop}
\end{equation}
is proportional to the relative production ratio and its dependence on the centre-of-mass energy is also reported here. This is measured using $\BsDsPi$ and $\BdDPi$ decays. Accurate knowledge of $f_{s}/f_{d}$ is a crucial input for every $\Bs$ branching fraction measurement, \eg $\BF(\Bs\rightarrow \mu^{+} \mu^{-})$, since it dominates in most cases the systematic uncertainty~\cite{LHCb-PAPER-2017-001}. Following the method described in Ref.~\cite{Fleischer:2010ca}, the value of $f_{s}/f_{d}$ can be calculated as
\begin{equation}
\dfrac{f_s}{f_d} = 0.982 \dfrac{\tau_{B_{d}}}{\tau_{B_s}} \dfrac{\mathcal{R}}{\mathcal{N}_{a}\mathcal{N}_{F}\mathcal{N}_{E}} \dfrac{\mathcal{B}(\Dm \rightarrow \Kp \pim \pim)}{\mathcal{B}(\Dsp \rightarrow \Kp \Km \pip)}, \label{eq:fsfd_abs}
\end{equation}
where $\mathcal{R}$ is defined in Eq.~\eqref{eq:fsfd_prop}, the numerical factor takes phase-space effects into account, $\mathcal{N}_{a}$ describes non-factorisable SU(3) breaking effects, $\mathcal{N}_{F}$ is the ratio of the form factors, $\mathcal{N}_{E}$ takes into account the contribution of the $W$-exchange diagram in the $\BdDPi$ decay,
and $\tau_{B_d}\ (\tau_{B_s})$ is the \Bd (\Bs) lifetime.

%% file: detector.tex
\section{Detector and simulation}
\label{sec:Detector}
The \lhcb detector~\cite{LHCb-DP-2008-001,LHCb-DP-2014-002} is a single-arm forward
spectrometer covering the \mbox{pseudorapidity} range $2<\eta <5$,
designed for the study of particles containing \bquark or \cquark
quarks. The detector includes a high-precision tracking system
consisting of a silicon-strip vertex detector surrounding the $pp$
interaction region~\cite{LHCb-DP-2014-001}, a large-area silicon-strip detector located
upstream of a dipole magnet with a bending power of about
$4{\mathrm{\,Tm}}$, and three stations of silicon-strip detectors and straw
drift tubes~\cite{LHCb-DP-2013-003,LHCb-DP-2017-001} placed downstream of the magnet.
The tracking system provides a measurement of the momentum, \ptot, of charged particles with
a relative uncertainty that varies from about 0.5\% below 20\gevc to 1.0\% at 200\gevc.
The minimum distance of a track to a primary vertex (PV), the impact parameter (IP), 
is measured with a resolution of $(15+29/\pt)\mum$,
where \pt is the component of the momentum transverse to the beam, in\,\gevc.
Different types of charged hadrons are distinguished using information
from two ring-imaging Cherenkov (RICH) detectors~\cite{LHCb-DP-2012-003}. 
Hadrons are identified by a calorimeter system consisting of
scintillating-pad and preshower detectors, an electromagnetic and a hadronic calorimeter. Muons are identified by a
system composed of alternating layers of iron and multiwire proportional chambers~\cite{LHCb-DP-2012-002}. \\
\indent The online event selection is performed by a trigger~\cite{LHCb-DP-2012-004}, 
which consists of a hardware stage, based on information from the calorimeter and muon
systems, followed by a software stage, which applies a full event
reconstruction. \\
\indent Simulation is required to calculate geometrical, reconstruction and selection efficiencies, and to determine shapes of invariant mass distributions.
In the simulation, $pp$ collisions are generated using
\pythia~\cite{Sjostrand:2007gs} with a specific \lhcb
configuration~\cite{LHCb-PROC-2010-056}.  Decays of unstable particles
are described by \evtgen~\cite{Lange:2001uf}, in which final-state
radiation is generated using \photos~\cite{Golonka:2005pn}. The
interaction of the generated particles with the detector, and its response,
are implemented using the \geant
toolkit~\cite{Allison:2006ve, *Agostinelli:2002hh} as described in
Ref.~\cite{LHCb-PROC-2011-006}.

%% file: selection.tex
\section{Selection }
\label{sec:Selection}
The $\BdDsPi$ ($\BdDPi$) decays are reconstructed by forming a \decay{\Dsp}{\Kp\Km\pip} (\decay{\Dm}{\Kp\pim\pim}) candidate and combining it with an additional pion of opposite charge, referred to as the \mbox{companion}. The same reconstruction and selection procedure is applied to the $\BsDsPi$ decay. For the $\BdDsPi$ decay, the invariant mass of the $\Kp\Km$ pair is required to be within $20\mevcc$ of the $\phi(1020)$ mass to select only the $\decay{\Dsp}{\phi(1020)\pip}$ decays, which significantly improves the signal-to-background ratio compared to other decays with a $\Kp\Km\pip$ combination in the final state. Selecting $\decay{\Dsp}{\phi(1020)\pip}$ decays has an efficiency of about $40\%$. \\
\indent At the hardware trigger stage, events are required to have a muon with high \pt or a hadron, photon or electron with high transverse energy in the calorimeters. For hadrons, the transverse-energy threshold varied between 3 and 4\gev between 2011 and 2016. The software trigger requires a two-, three- or four-track secondary vertex with significant displacement from any primary $pp$ interaction vertex (PV). At least one charged particle must have transverse momentum $\pt > 1.6\gevc$ and be inconsistent with originating from a PV. A multivariate algorithm~\cite{BBDT} is used for the identification of secondary vertices consistent with the decay of a \bquark hadron. \\
\indent After the trigger selection, a preselection is applied to the reconstructed candidates to ensure good quality for the vertex of the \bquark-hadron and \cquark-hadron candidates comprising of tracks with large total and transverse momentum. Combinatorial background is suppressed using a gradient boosted decision tree (BDTG) algorithm~\cite{Breiman,Roe:2004na}, trained on \mbox{Run 1} $\BsDsPi$ data. A set of 15 variables is used to train the BDTG classifier, the ones with highest importance in the training being the transverse momentum of the companion pion, the radial flight distance of the $\Bsb$ and of the $\Dsp$ candidates, the minimum transverse momentum of the $\Dsp$ decay products and the minimum \chisqip of the companion and the $\Bsb$ candidates, where \chisqip is defined as the difference in the vertex-fit $\chi^2$ of a given PV reconstructed with and without the particle under consideration. The correlation among the input variables has been studied and was found to be small. The BDTG classifier used in this measurement is described in  Ref.~\cite{Eitschberger:2018ofp}. \\
\indent To improve the \Bd and \Bsb invariant mass resolutions, the \Dsp and \Dm invariant masses are constrained to their known values~\cite{PDG2020}. All $\Dsp\pim$ ($\Dm\pip$) candidates are required to have their invariant masses, $m(\Dsp\pim)$ ($m(\Dm\pip)$), within the range $5150\text{--}5800$ $(5000\text{--}5800) \mevcc$ and the $\Kp\Km\pip$ ($\Kp\pim\pim$) invariant mass within $1930\text{--}2065$ $(1830\text{--}1920) \mevcc$. The range of the $\Kp\Km\pip$ invariant mass includes a large upper sideband to model properly the combinatorial background shape, as described in Sec.~\ref{section:parametrisations}.\\
\indent To reduce the background due to misidentified final-state particles, particle identification (PID) information from the RICH detectors is used.
The companion pion is required to pass a strict PID requirement to reduce the number of $\BsDsK$ ($\BdDK$) decays where the kaon companion is misidentified as a pion. 
For $\decay{\Dsp}{\phi(1020)\pip}$ candidates, loose PID requirements are applied to both kaons and the pion, which imply a signal efficiency of about $96\%$. In the case of the pion, the PID requirement is used primarily to remove protons originating from the $\Lc \to \phi p$ decay.
Further PID requirements are applied to veto \mbox{$\decay{\Lb}{\Lc(\rightarrow p\Km\pip)}\pim$} and \mbox{$\decay{\Bdb}{\Dp(\rightarrow \Km\pip\pip)\pim}$} and $\decay{\Lbbar}{\Lcbar(\rightarrow \antiproton\Kp\pim)}\pip$ and $\decay{\Bs} {\Dsm (\rightarrow \Km\Kp\pim)\pip}$ events, which are misidentified as the final-state particles of $\Dsp(\rightarrow\Kp\Km\pip)\pim$ and $\Dm(\rightarrow\Kp\pim\pim)\pip$ decays, respectively. These vetoes are applied if candidates are consistent with the above mentioned decays when a mass hypothesis is changed. The PID requirements result in $75\%$ efficiency for \BdDsPi signal decays, which is dominated by the strict PID requirement on the companion pion, while the retention is about $9\%$ for the \BsDsK misidentified background contribution. \\
\indent The event selection efficiencies are calculated from simulation with the exception of the efficiency of the PID requirements which is determined using calibration data samples.

%% file: parametrisations.tex
\section{Signal and background parametrisation} \label{section:parametrisations}

After the full event selection, unbinned maximum-likelihood fits are performed to obtain the yields of the signal $\BdDsPi$ and the normalisation $\BdDPi$ candidates. A two-dimensional fit to the $\Dsp\pim$ and the $\Kp\Km\pip$ invariant mass distributions is performed to determine the $\BdDsPi$ signal yield, while the yield of the normalisation channel is obtained from a fit to the $\Dm\pip$ invariant mass distribution. Due to the $\Dsp$ mass constraint, the correlation between $m(\Dsp\pim)$ and $m(\Kp\Km\pip)$ is found to be small, thus the two variables are factorised in the fit model~\cite{LHCb-PAPER-2017-047}. The two-dimensional fit is performed in order to constrain the combinatorial background (see further in this Section for details). \\
\indent The $\BdDsPi$ decay is Cabibbo-suppressed and is therefore considerably less abundant than the Cabibbo-favoured $\BsDsPi$ decay, which produces the same final-state particles. The $m(\Dsp\pim)$ and $m(\Dm\pip)$ shapes for $\BsDsPi$ and $\BdDPi$ candidates, respectively, are described by the sum of a double-sided Hypatia function~\cite{Santos:2013gra} and a \mbox{Johnson $S_U$} function~\cite{johnson1949systems}. The left tail of the $\BsDsPi$ invariant mass distribution overlaps with the $\BdDsPi$ signal peak and therefore special attention is given to the description of the lower mass range of the $\BsDsPi$ peak, shaped by the combination of detector resolution and radiative effects. The $\BdDsPi$ signal is described with the same model as the $\BsDsPi$ decay, shifted by the known $\Bd$--$\Bs$ mass difference~\cite{PDG2020}. The left tail of this distribution is described by two parameters, $a_{1}$ and $n_{1}$, which are found to be correlated and therefore the parameter $n_{1}$ is fixed to the value obtained from simulation, whereas $a_{1}$ is obtained from simulated $\BsDsPi$ and $\BdDPi$ events, as well as from $\BdDPi$ data. In the invariant mass fit to $\BdDPi$ candidates the common mean of the double-sided Hypatia and the Johnson $S_U$ functions, the widths and the left-tail parameter $a_{1}$ are left free in the fit, while this parameter is constrained in
the $\Dsp\pim$ invariant mass distribution, as the background does not allow to determine the shape of the radiative tail reliably. All other parameters are fixed from simulation.  
In the $\Kp\Km\pip$ invariant mass fit a sum of two Crystal Ball functions with a common mean is used. The common mean and a scale factor for the widths are left free, while the other shape parameters are fixed from simulation. \\
\indent The combinatorial background in  \mbox{$\BdDsPi$} candidates is split in two components, referred to as random-\Dsp and true-\Dsp. The random-\Dsp combinatorial background consists of random combinations of tracks that do not peak in the $\Kp\Km\pip$ invariant mass, while the true-\Dsp combinatorial background consists of events with a true \Dsp meson, combined with a random companion track. The upper mass range of the $\Kp\Km\pip$ candidate sample is used to account accurately for the random-$\Dsp$ component, modelled with a single exponential distribution, while the true-$\Dsp$ background is described by the signal shape. In the \Dsp\pim invariant mass fit, the random-$\Dsp$ background is described by an exponential distribution and the true-$\Dsp$ background is described by the sum of an exponential and a constant function. The exponential parameters are left free in both invariant mass fits.  \\
\indent The combinatorial background in the $m(\Dm\pip)$ fit of the normalisation channel is described by the sum of an exponential and a constant function, with the relative weight of the two functions and exponential parameter left free.\\
\indent Decays where one or more final-state particles are not reconstructed are referred to as partially reconstructed backgrounds. In the $\Dsp\pim$ and $\Dm\pip$ invariant mass fits these background contributions are described by an upward-open parabola or a parabola exhibiting a maximum, whose ranges are defined by the kinematic endpoints of the decay, which are convolved with Gaussian resolution functions, and which are known to describe decays involving a missing neutral pion or a missing photon, as defined in Ref.~\cite{LHCb-PAPER-2017-021}. In the fit to the $\Kp\Km\pip$ invariant mass, the partially reconstructed background contributions are described by the signal mass shape. \\
\indent The $m(\Dsp\pim)$ fit requires two partially reconstructed background components from \mbox{$\decay{\Bsb}{\Dssp(\rightarrow \Dsp \gamma/\piz)\pim}$} and $\decay{\Bsb}{\Dsp\rhom(\rightarrow\pim\piz)}$ decays. The fit model describing the $\Dm\pip$ invariant mass accounts analogously for two partially reconstructed background contributions: $\decay{\Bd}{\Dstarm(\rightarrow \Dm\piz)\pip}$ and $\decay{\Bd}{\Dm\rhop(\rightarrow \pip \piz)}$. In the case of the $\BsDsstPi$ background the previously mentioned upward-open parabola together with a parabola exhibiting a maximum is used to parameterise the components with $\Dssp\rightarrow\Dsp\gamma$ and $\Dssp\rightarrow\Dsp\piz$ decays, respectively. The $\BsDsRho$ background is described by the upward-open parabola, to take into account the missing neutral pion. The $\BdDstPi$ decay uses an upward-open parabola function and exhibits a double-peaked shape. Most parameters are obtained from simulated events and fixed, aside from the relevant invariant mass shifts and widths. For the $\BdDRho$ background a single upward-open parabola function is taken, with a floating width and a floating mass shift parameter that is shared with the $\BdDstPi$ contribution. The widths of the partially reconstructed background contributions in the $m(\Dsp\pim)$ fits are fixed to the values obtained from $\BdDPi$ candidates in data, corrected for differences between the $m(\Dsp\pim)$ and $m(\Dm\pip)$ distributions, as obtained from simulation.\\
\indent The $\BdDPi$ candidate sample is contaminated by the $\Bs \rightarrow \Dsm\pip$, \mbox{$\Lbbar \rightarrow \Lcbar \pip$} and $\BdDK$ decays, resulting from the misidentification of one or two of the final-state particles. Analogously, the $\BsDsK$, $\LbLcPi$ and $\Bdb \rightarrow \Dp \pim$ decays are misidentified background contributions of the $\BdDsPi$ candidate sample. Their shapes are determined from simulation using a non-parametric kernel estimation method~\cite{Cranmer:2000du}. The yields of the misidentified background contributions are estimated by using known branching fractions~\cite{PDG2020} and efficiencies that are determined from simulated background decays. Each yield of a misidentified background in the fit model is constrained to be close to its estimated value and is allowed to vary within the corresponding uncertainty.

%% file: massFits.tex
\section{Signal yields} \label{sec:massFits}
\begin{figure}
\centering
\includegraphics[width=0.495\textwidth]{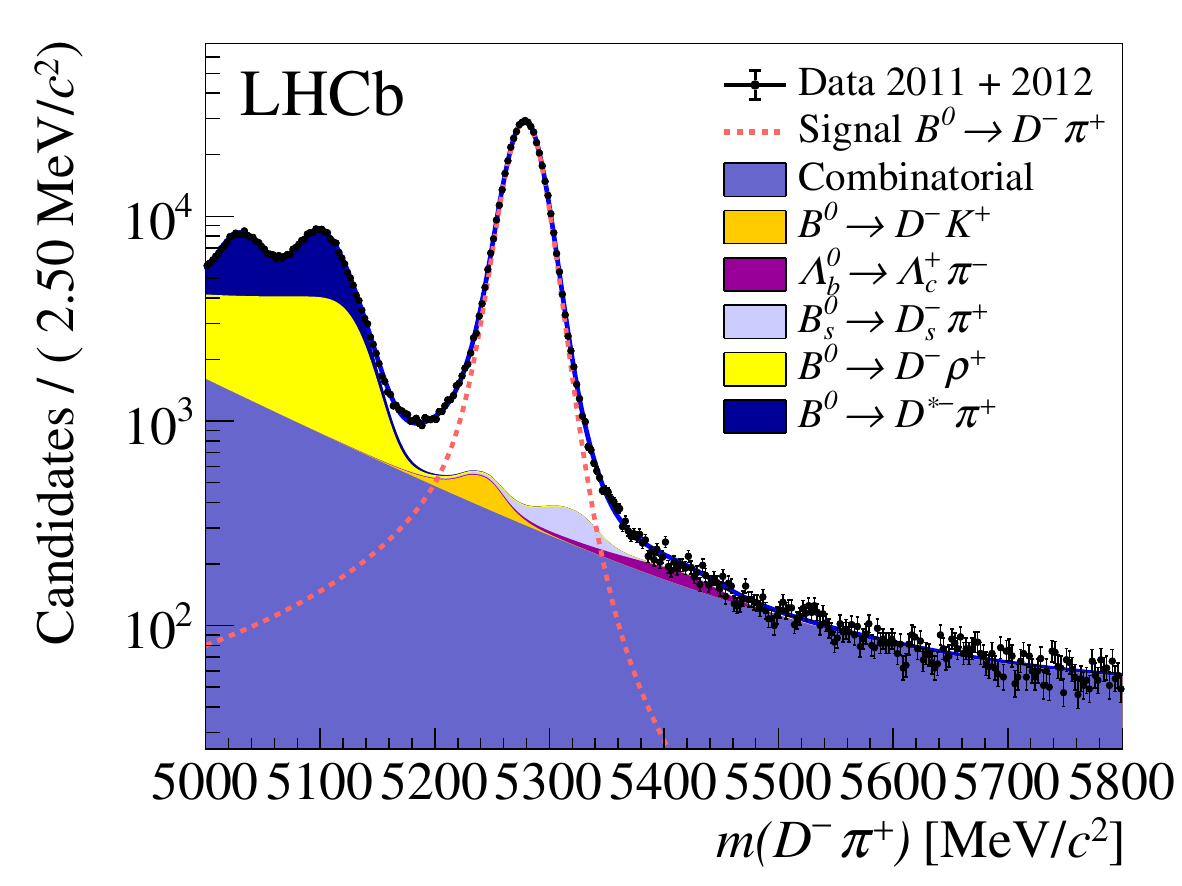}
\includegraphics[width=0.495\textwidth]{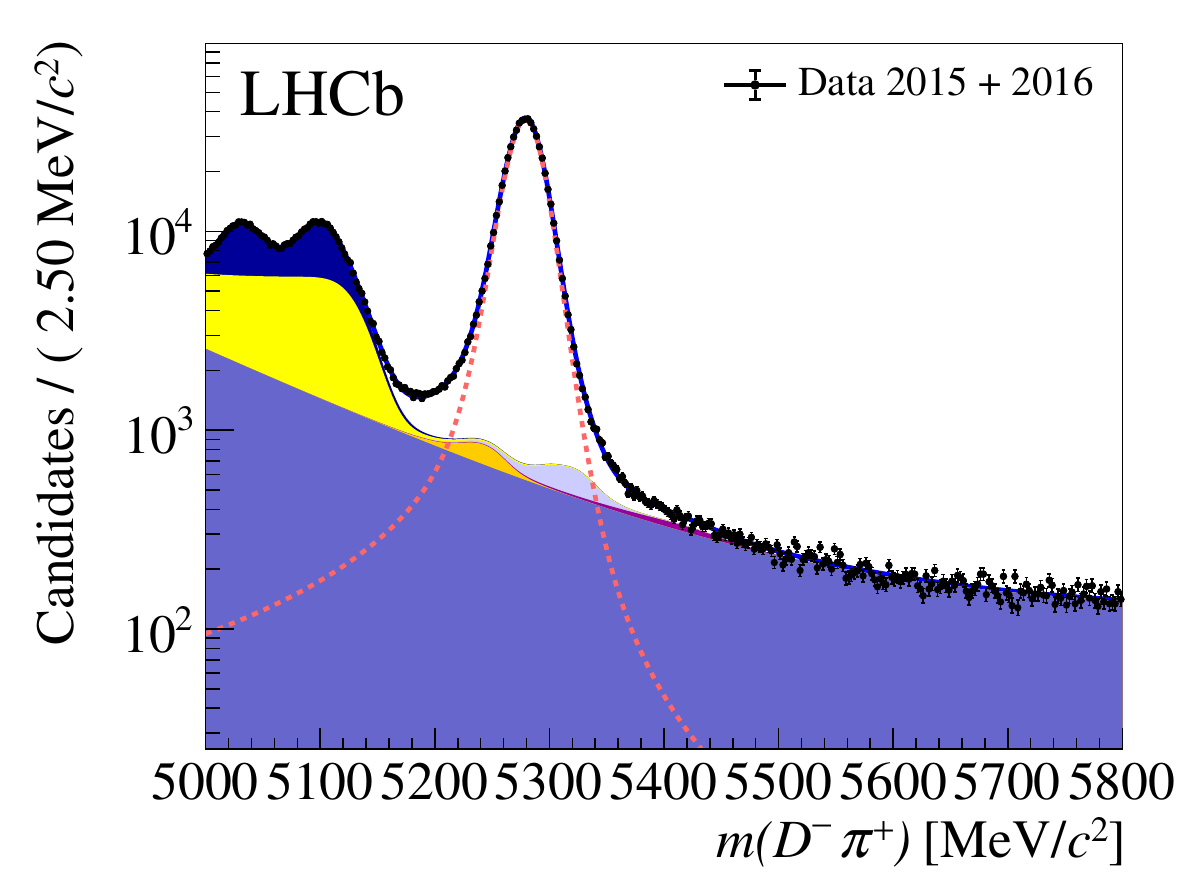}
\caption{The invariant mass distributions of normalisation \BdDPi candidates, for (left) Run~1 and (right) Run~2 data samples. Overlaid are the fit projections along with the signal and background contributions.}
\label{fig:DataFit_Bd2DPi}
\end{figure}

\begin{figure}
\centering
\includegraphics[width=0.495\textwidth]{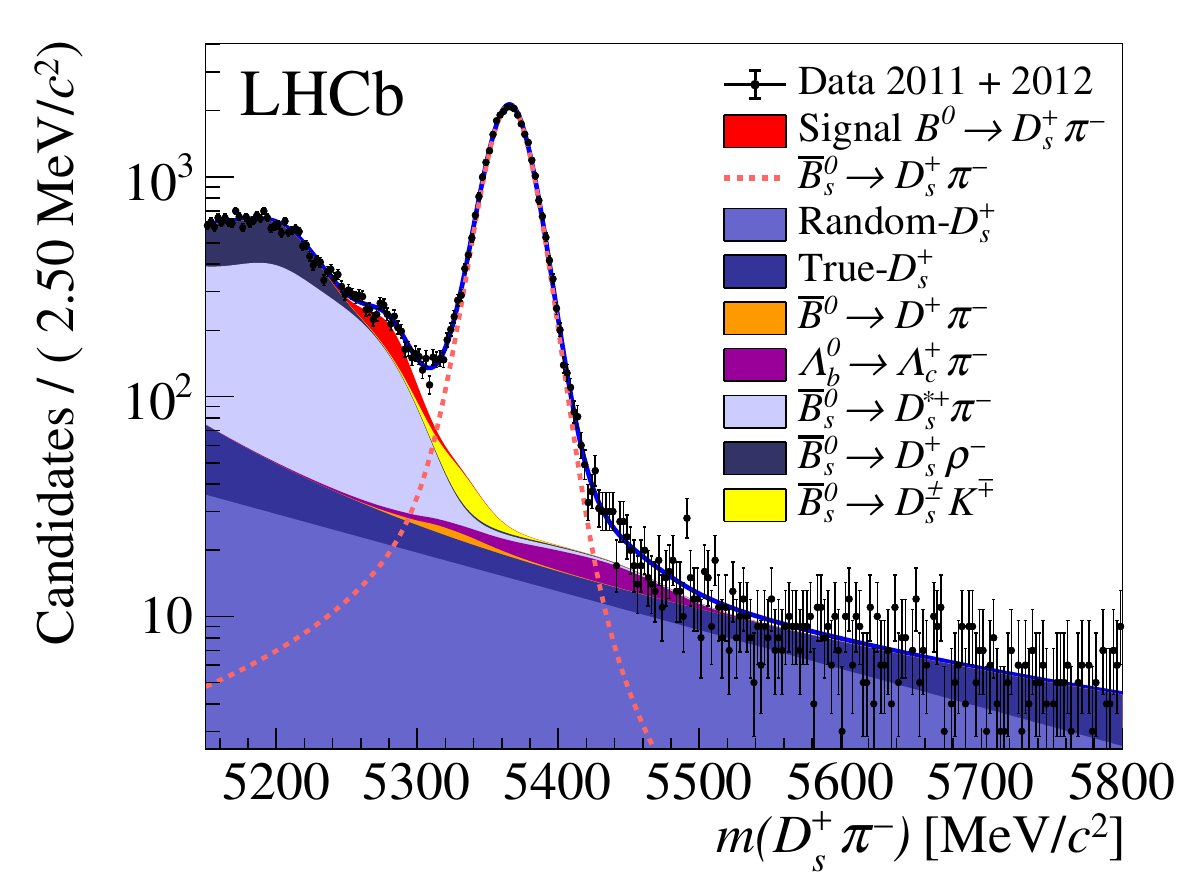}
\includegraphics[width=0.495\textwidth]{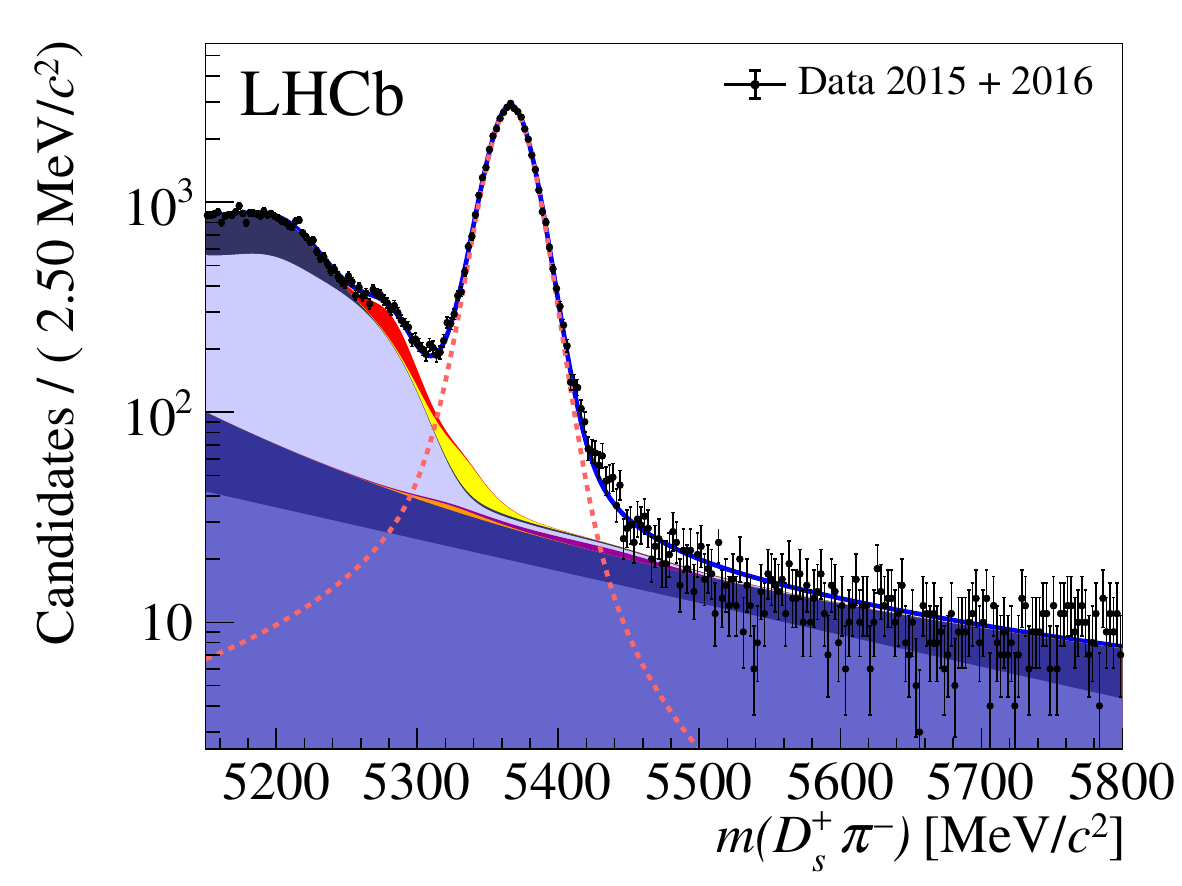}\\
\vspace{+0.2cm}
\includegraphics[width=0.495\textwidth]{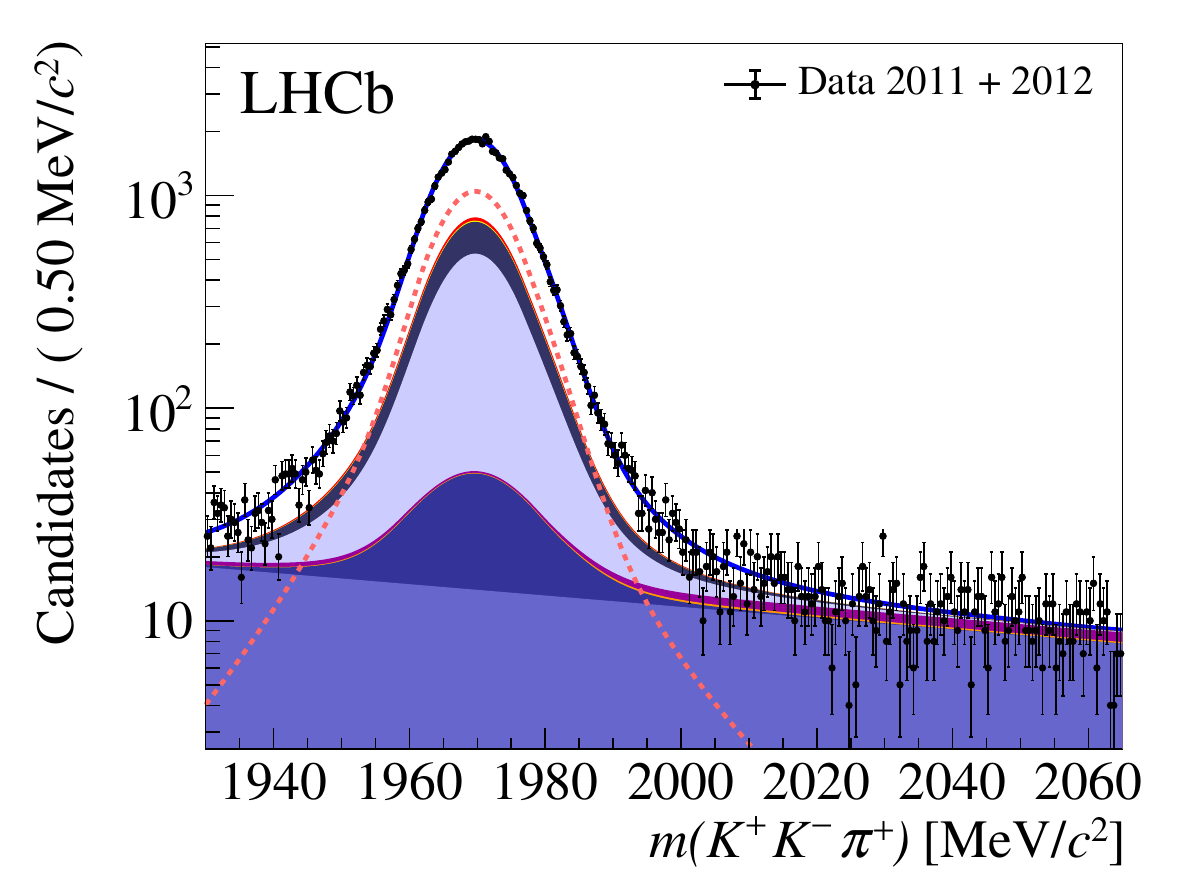}
\includegraphics[width=0.495\textwidth]{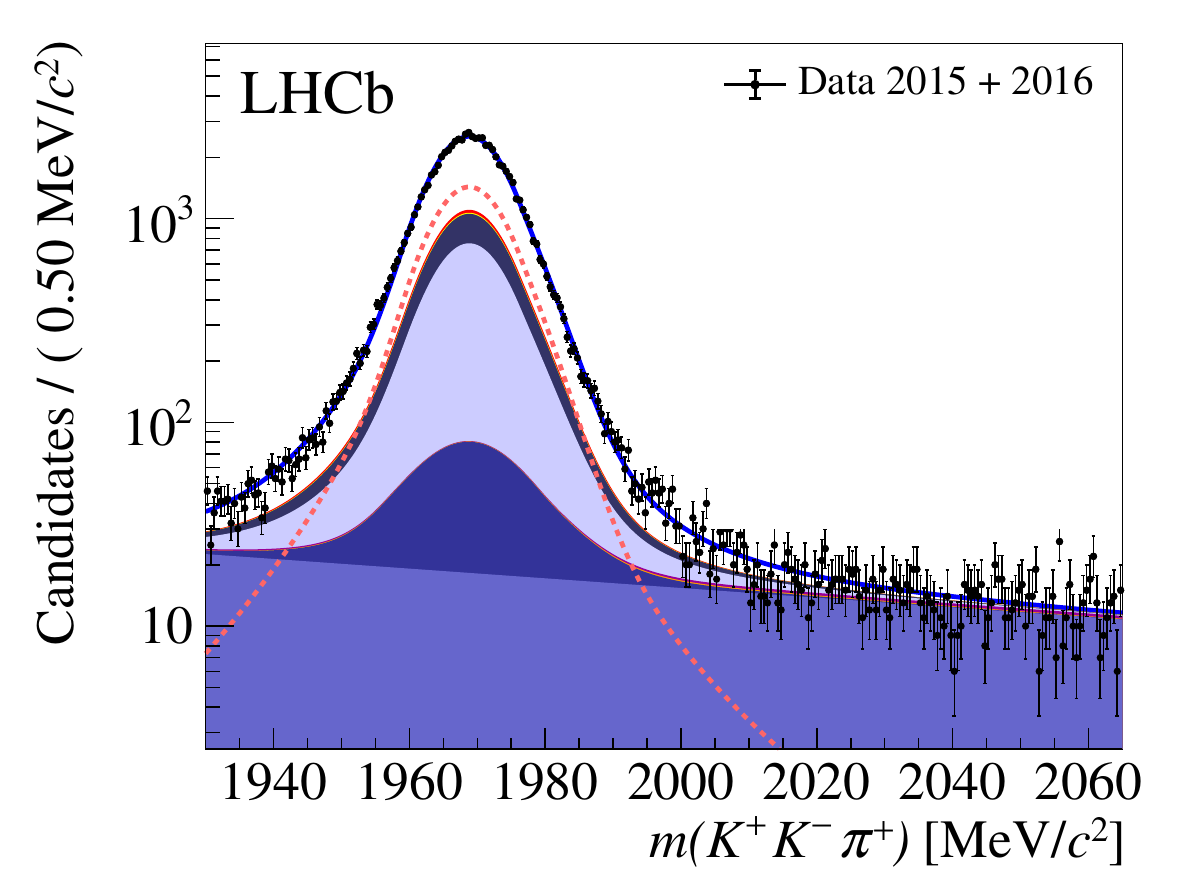}
\caption{The (top) $\Dsp\pim$ and (bottom) $\Kp\Km\pip$ invariant mass distributions of signal \BdDsPi candidates, for (left) Run~1 and (right) Run~2 data samples. Overlaid are the fit projections along with the signal and background contributions.}
\label{fig:DataFit_BsDsPi}
\end{figure}

The $m(\Dm\pip)$ data distributions, with overlaid fit projections for the total, the $\BdDPi$ signal and the background components, are shown in Fig.~\ref{fig:DataFit_Bd2DPi}. The resulting signal yields are $\YieldBdDPiRunOnePaper$ and $\YieldBdDPiRunTwoPaper$ for Run 1 and Run 2 samples, respectively. 
The fit results are also used to constrain the left tail of the signal shape and the widths of the partially reconstructed backgrounds to the invariant mass distribution of $\BdDsPi$ candidates. \\
\indent The two-dimensional fit to $\BdDsPi$ candidates is performed in the $\Dsp\pim$ and $\Kp\Km\pip$ invariant mass distributions. The $\BdDsPi$ branching fraction is determined using the yields of the signal and normalisation modes, their selection efficiencies and the known $\BdDPi$, $\DKPiPi$ and $\DsKKPi$ branching fractions~\cite{PDG2020}. The two-dimensional fit is performed simultaneously for Run~1 and Run~2 data samples in which the $\BR(\BdDsPi)$ and left-tail parameter are shared. The fit results in $\BdDsPi$ signal yields of $\YieldBdDsPiRunOnePaper$ and $\YieldBdDsPiRunTwoPaper$ and $\BsDsPi$ yields of $\YieldBsDsPiRunOnePaper$ and $\YieldBsDsPiRunTwoPaper$ for Run 1 and Run 2 samples, respectively. Figure~\ref{fig:DataFit_BsDsPi} shows the $\Dsp\pim$ invariant mass distributions together with the fit projections and background contributions overlaid. Additionally, the invariant mass fits to $\BdDPi$ and $\BdDsPi$ candidates are performed simultaneously to 2011, 2012 and Run~2 data in order to study the collision energy dependence of $f_{s}/f_{d}$, as is described in Sec.~\ref{sec:Resuls}.

%% file: systematics.tex
\section{Systematic uncertainties}
\label{sec:Systematics}
Systematic uncertainties on the $\BF(\BdDsPi)$ measurement arise from choices in the fit model and the determination of trigger, BDT and PID efficiencies. Many possible sources of systematic uncertainty cancel in the ratio of either the yields or the efficiencies of $\BdDsPi$ and $\BdDPi$ events. A summary of all the systematic uncertainties is shown in Table~\ref{tab:syst_uncertainty}. The precision of the measurement relies mostly on the accurate modelling of the signal shape and of the partially reconstructed backgrounds.\\
\indent The most critical aspect of the signal shape is the description of the left tail of the \BsDsPi signal, affecting the composition of signal and background around the \Bd mass. The shape of the left tail was determined from \BdDPi candidates, taking into account differences between the final states, as obtained from simulation, and was Gaussian constrained in the fit. A systematic uncertainty is assigned for the assumption of the signal shape. This is done by repeating the signal fit with a different parametrisation, \ie the sum of a double-sided Hypatia function and a Gaussian function, which leads to a systematic uncertainty of $5.1\%$. This parametrisation was found to be the only alternative parametrisation that satisfactorily described simulated signal candidates. Furthermore, a systematic uncertainty is assigned by fixing the mean of the $\BdDsPi$ signal shape to the result of the $\BdDPi$ fit, rather than shifting by the known $\Bd$--$\Bs$ mass difference. Moreover, the width of the $\BdDsPi$ signal shape is scaled by the ratio of the known $\Bd$ and $\Bs$ masses.
The widths of the partially reconstructed backgrounds is varied by $\pm 1 \mevcc$, in order to cover the differences between data and simulation as well as the differences between the $\Dsp\pim$ and $\Dm\pip$ invariant mass distributions. The resulting difference between the signal yields is assigned as a systematic uncertainty.\\ 
\indent The simulated samples are corrected for an imperfect modelling of the response of the particle identification algorithms as a function of the kinematical properties of the particle, using samples of $\Dstarp$ calibration data. A systematic uncertainty associated with the PID efficiency evaluation is assigned by varying the corrections within their uncertainties. Proton misidentification is the most difficult to control accurately from data calibration samples, as relatively little calibration data is available in the kinematic region that overlaps with the $B$ decay products. In addition, the Cherenkov angles of photons emitted by protons and kaons are more similar than those of kaons and pions. Thus, a systematic uncertainty is  estimated from the difference between the nominal signal yields and a fit where the misidentified background $\LbLcPi$ decay yield is left free to vary. \\
\indent The systematic uncertainty assigned to the hardware trigger efficiency takes into account a difference in detection efficiency between kaons and pions. This mostly cancels in the ratio of $\BdDPi$ and $\BdDsPi$ efficiencies, but the difference of one final-state particle is sensitive to this detection asymmetry. Moreover, an uncertainty related to the reconstruction efficiency of charged particles is taken into account, which mainly arises from the uncertainty on the LHCb material and the different interaction cross-section of pions and kaons with the material~\cite{LHCb-PAPER-2018-036}. Additionally, a systematic uncertainty is determined on the BDT efficiency due to the difference between simulation and data. This is determined by weighting all the BDT input variables in the simulated signal sample to the signal distributions in data, which are obtained using signal weights for each candidate using the \textit{sPlot} technique~\cite{Pivk:2004ty}. 

\begin{table}
\centering
\caption{Relative systematic uncertainty $\sigma$ on the \mbox{$\BdDsPi$} branching fraction measurement.}
\begin{tabular}{lc}
\hline \hline
Source                                              & $\sigma(\BF(\BdDsPi)$) [\%] \\
\hline
Fit model\\
\hspace{0.25cm} Signal shape parametrisation        & $5.1$ \\
\hspace{0.25cm} \BdDsPi signal width                & $1.5$ \\
\hspace{0.25cm} \BdDsPi mean                        & $0.2$ \\
\hspace{0.25cm} Partially reconstructed backgrounds & $4.2$ \\
\hspace{0.25cm} Misidentified backgrounds                   & $0.6$ \\
Efficiencies\\
\hspace{0.25cm} Hardware trigger efficiency         & $0.3$ \\
\hspace{0.25cm} Reconstruction efficiency           & $0.5$ \\
\hspace{0.25cm} BDT efficiency                      & $0.7$ \\
\hspace{0.25cm} PID efficiency                      & $1.1$ \\
\hline
Total                                               & $6.9$ \\
\hline \hline
\end{tabular}
\label{tab:syst_uncertainty}
\end{table}

\indent The systematic uncertainties on the collision energy dependence of the efficiency-corrected $\BsDsPi$ and $\BdDPi$ yield ratios are shown in Table~\ref{tab:syst_fdfd}. The sources of these systematic uncertainties are the same as for the $\BdDsPi$ branching fraction. Exceptions are the uncertainties on the $\BdDsPi$ signal and the partially reconstructed backgrounds, which are found to be negligible, and the uncertainty on the charged-particle reconstruction efficiency, which cancels out in the double ratio of efficiencies.
\begin{table}
\centering
\caption{Relative systematic uncertainty $\sigma$ on the ratio of the efficiency-corrected $\BsDsPi$ and $\BdDPi$ yield ratios. The ratios $\mathcal{R}_{13\tev}/\mathcal{R}_{7\tev}$ and $\mathcal{R}_{13\tev}/\mathcal{R}_{8\tev}$ are reported together as the difference of the systematic uncertainty for 7 and 8 \tev is negligible.}
\begin{tabular}{lccc}
\hline \hline
Source                                              & \multicolumn{2}{c}{$\sigma\left(\dfrac{\mathcal{R}_{13\tev}}{{\mathcal{R}_{7,8\tev}}}\right)$} [\%] & $\sigma\left(\dfrac{\mathcal{R}_{8\tev}}{\mathcal{R}_{7\tev}}\right)$ [\%] \\
\hline
Fit model\\
\hspace{0.25cm} Signal shape parametrisation        & \multicolumn{2}{c}{$0.2$} & -- \\
\hspace{0.25cm} Misidentified backgrounds                   & \multicolumn{2}{c}{$0.2$} & -- \\
Efficiencies\\
\hspace{0.25cm} Hardware trigger efficiency         & \multicolumn{2}{c}{$0.4$} & $0.4$ \\
\hspace{0.25cm} BDT efficiency                      & \multicolumn{2}{c}{$1.1$} & $1.3$\\
\hspace{0.25cm} PID efficiency                      & \multicolumn{2}{c}{$1.4$} & $1.4$\\
\hline
Total                                               & \multicolumn{2}{c}{$1.9$} & $2.0$\\
\hline \hline
\end{tabular}
\label{tab:syst_fdfd}
\end{table}

%% file: results.tex
\section{Results}
\label{sec:Resuls}
\begin{figure}
\centering
\vspace*{0cm}
\includegraphics[scale=0.6]{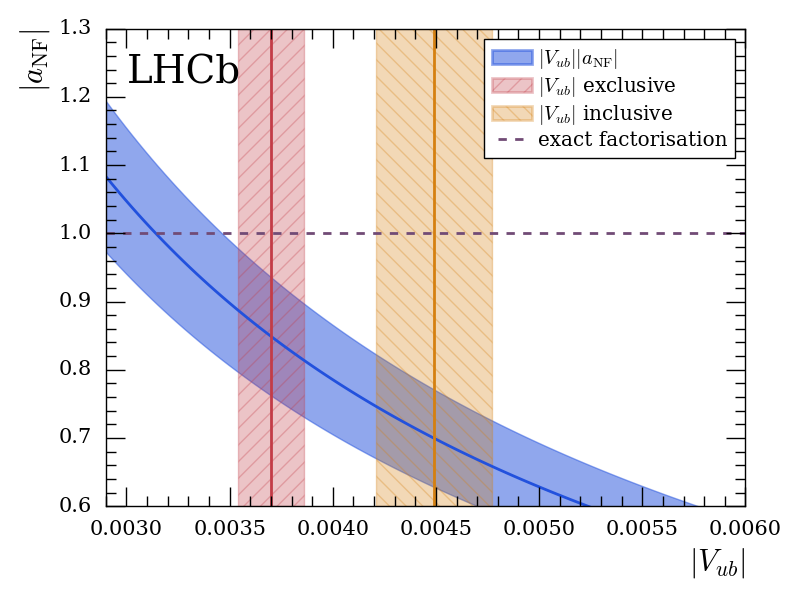}
\caption{Result of the determination of $|V_{ub}||a_{\text{NF}}|$. The blue line represents the result of this measurement, the vertical bands are the known exclusive and inclusive measurements of $|V_{ub}|$, which are $(3.70\pm 0.16)\times 10^{-3}$ and $(4.49\pm 0.28)\times 10^{-3}$, respectively~\cite{PDG2020}. The horizontal dashed line at $|\aNF|=1.0$ represents exact factorisation. The error bands represent an uncertainty of one standard deviation.} 
\label{fig:resultPlot}
\end{figure}

\begin{figure}
\centering
\includegraphics[scale=0.5]{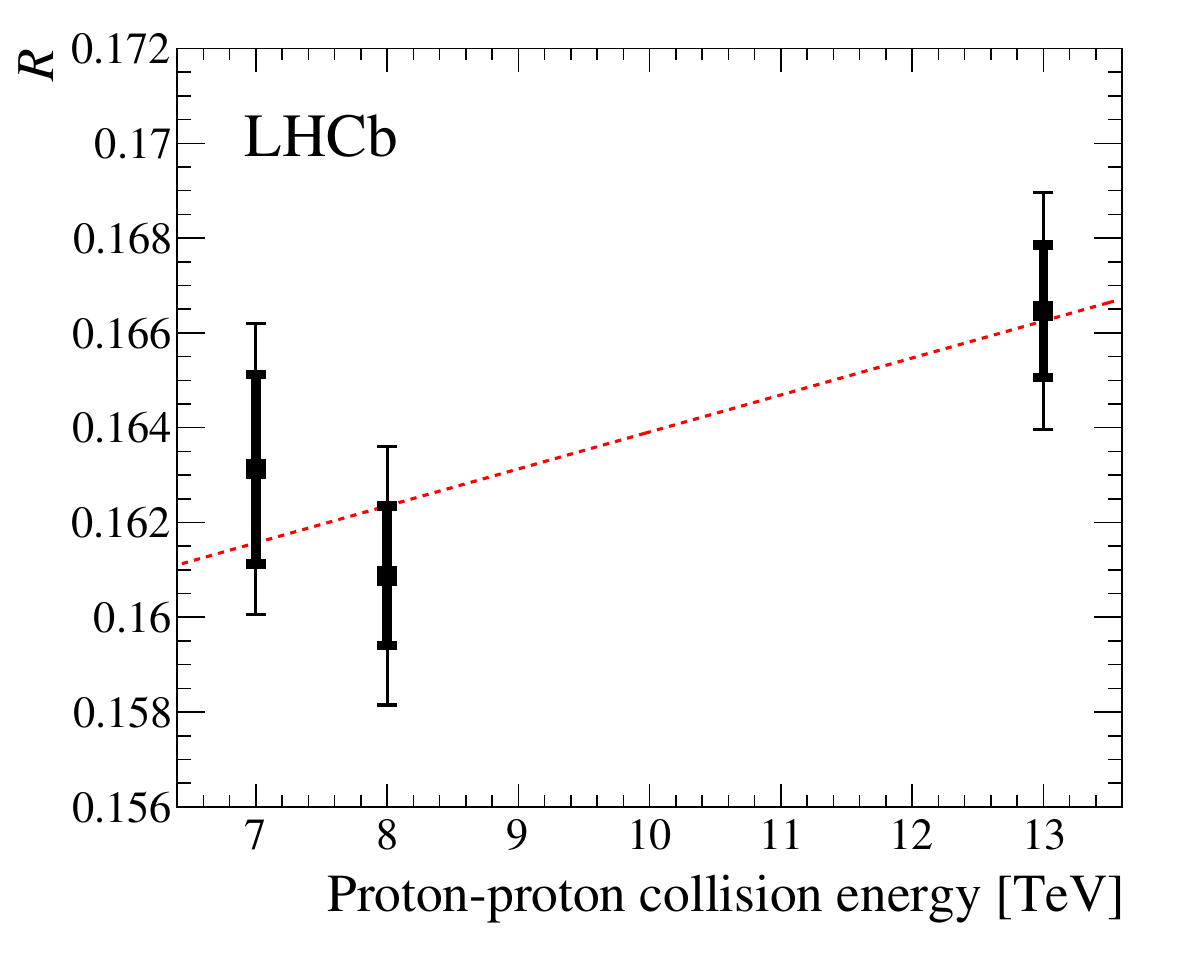}
\caption{Visualisation of the $pp$ collision energy dependence of the efficiency-corrected yield ratio of $\BsDsPi$ and $\BdDPi$ decays, which scales with $f_{s}/f_{d}$. The inner error bars indicate the statistical uncertainty only, whereas the outer indicate the uncorrelated, including statistical, uncertainties. The correlated systematic uncertainty is not shown. The red dotted line represents a linear fit through the three values of $\mathcal{R}$ with uncorrelated, including statistical, uncertainties.}
\label{fig:R_energyDependence}
\end{figure}

Table~\ref{tab:results} gathers all measurements and inputs to determine the branching fraction according to Eq.~\eqref{eq:BF_Bd2DsPi}. 
\begin{table}
\centering
\caption{Results of $\BdDsPi$ and $\BdDPi$ signal efficiencies and yields, as well as the branching fractions used as input for this measurement~\cite{PDG2020}.}
\vspace{-2mm}
\begin{tabular}{lcc}
\hline \hline
                            & Run 1               & Run 2               \\
                            \hline
$\epsilon_{\BdDsPi}$ ($\%$) & $\EffBdDsPiRunOne$\hspace{0.3cm}  & $\EffBdDsPiRunTwo$\hspace{0.3cm} \\
$\epsilon_{\BdDPi}$ ($\%$)  & $\EffBdDPiRunOne$\hspace{0.3cm}   & $\EffBdDPiRunTwo$\hspace{0.3cm}  \\
$N_{\BdDPi}$                & \hspace{0.8cm}$\YieldBdDPiRunOnePaper$  & \hspace{0.8cm}$\YieldBdDPiRunTwoPaper$  \\
$N_{\BdDsPi}$               & \hspace{0.8cm}$\YieldBdDsPiRunOnePaper$ & \hspace{0.8cm}$\YieldBdDsPiRunTwoPaper$ \\
\cmidrule(lr){1-3}
$\BF(\BdDPi)$               & \multicolumn{2}{c}{$(2.52 \pm 0.13) \times 10^{-3}$} \\
$\BF(\DKPiPi)$              & \multicolumn{2}{c}{$(9.38 \pm 0.16) \times 10^{-2}$} \\
$\BF(\DsKKPi) $             & \multicolumn{2}{c}{$(5.39 \pm 0.15) \times 10^{-2}$} \\
\hline \hline
\end{tabular}
\label{tab:results}
\end{table}
The branching fraction ratio of \BdDsPi and \BdDPi decays is found to be
\begin{equation*}
    \frac{\BF(\BdDsPi)}{\BF(\BdDPi)} = \RatioBFBdDsPiBFBdDPiPaper,
    \label{BF_ratio}
\end{equation*}
where the first uncertainty is statistical, the second systematic and the third stems from knowledge of the $\Dm\rightarrow\Kp\pim\pim$ and $\Dsm\rightarrow\Km\Kp\pim$ branching fractions.\\
\indent Using the known value of $\BF(\BdDPi)$~\cite{PDG2020}, the \BdDsPi branching fraction is found to be
\begin{equation*}
    \BF(\BdDsPi) = \BFBdDsPiPaper, \label{eq:combinedBF}
\end{equation*}
where the first uncertainty is statistical, the second systematic and the third refers to the uncertainty due to the branching fractions listed in Table~\ref{tab:results}. This result represents the most precise single measurement of $\BF(\BdDsPi)$ to date. \\
\indent The $\BdDsPi$ branching fraction depends on both $|\aNF|$ and $|\Vub|$. Using the measurement of $\BF(\BdDsPi)$, the product
\begin{equation*}
    |V_{ub}| |\aNF| = \VubaNFExpanded
\end{equation*}
is obtained, where the first uncertainty is from the $\BdDsPi$ branching fraction measurement and the second from the CKM and QCD parameters. The form factor $F(\Bd\rightarrow\pim)|_{q^2=m^2_{\Dsp}} = 0.327\pm0.025$ is obtained using light-cone sum rules~\cite{formfactor,Ball:2006jz} and lattice QCD calculations are used for the decay constant $f_{\Dsp} = 0.2499 \pm 0.0005 \gev$~\cite{decayconstant,decayconstant2}. A phase-space factor $\Phi = 296.2\pm0.8\gev^{-2}$ is used in order to relate the branching fraction to $|\Vub||\aNF|$. Additionally, the CKM matrix element $|V_{cs}|$ is well measured and used as an input~\cite{PDG2020}. The determination of $|\Vub||\aNF|$ can be compared to the known inclusive and exclusive determinations of $|\Vub|$ to provide a constraint on the $|\aNF|$ parameter as displayed in Fig.~\ref{fig:resultPlot}. \\
\indent The branching fraction ratio of $\BdDsPi$ and $\BdDPi$ decays can be used to determine the parameter $r_{D\pi}$, as shown in Eq.~\eqref{eq:rDPi}. Inserting the measured branching fraction ratio $\BF(\BdDsPi)/\BF(\BdDPi)$, the tangent of $\theta_{c}$~\cite{PDG2020} and the fraction between the decay constants $f_{\Dsp}$ and $f_{\Dp}$~\cite{decayconstant,decayconstant2} into Eq.~\eqref{eq:rDPi} gives
\begin{equation*}
    r_{D\pi} = \rDPiExpanded,
\label{eq:rDPI_value}
\end{equation*}
where the first uncertainty is statistical, the second systematic and the third arises from possible non-factorisable \grpsuthree-breaking effects, estimated to be $20\%$ according to Ref.~\cite{DeBruyn:2012jp}. \grpsuthree-breaking effects of about $20\%$ are consistent with the measured $|\aNF|$ in this analysis, see Fig.~\ref{fig:resultPlot}. \\
\indent Finally, the potential dependence of the hadronisation fraction $f_{s}/f_{d}$ on collision energy is probed using the $\BdDPi$ and $\BsDsPi$ signal yields obtained in the invariant mass fits, using Eq.~\eqref{eq:fsfd_prop}. To determine these, the fit to Run~1 data is split based on collision energy into 2011 ($7\tev$) and 2012 ($8\tev$), sharing the shape parameters. 
The measured double ratios for the different collision energies are
\begin{align*}
    \mathcal{R}_{13\tev}/\mathcal{R}_{7\tev} &= \fsfdThirteenSevenPaper, \\
    \mathcal{R}_{13\tev}/\mathcal{R}_{8\tev} &= \fsfdThirteenEightPaper,\\
    \mathcal{R}_{8\tev}/\mathcal{R}_{7\tev} &= \fsfdEightSevenPaper,
\end{align*}
where the first uncertainty is statistical and the second systematic. The average transverse momentum of the $B$ meson after full event selection is found to be $10.4$, $10.6$ and $10.9\gevc$ for $pp$ collision centre-of-mass energies of $7\tev$, $8\tev$ and $13\tev$, respectively. The separate values of $\mathcal{R}$ at the three collision energies are
\begin{align*}
    \mathcal{R}_{7\tev} &= \RSevenPaper,\\
    \mathcal{R}_{8\tev} &= \REightPaper,\\
    \mathcal{R}_{13\tev} &= \RThirteenPaper,
\end{align*}
where the first uncertainty is statistical and the following are the uncorrelated and correlated systematic uncertainties, respectively. The value of $\mathcal{R}$ at $7\tev$ shows good agreement with the previous hadronic $f_{s}/f_{d}$ measurement at $7\tev$, which was performed using $\BsDsPi$, $\BdDPi$ and $\BdDK$ decays~\cite{LHCb-PAPER-2012-037}. A visualisation of the dependence of $\mathcal{R}$ on the centre-of-mass energy is given in Fig.~\ref{fig:R_energyDependence}. The resulting centre-of-mass energy dependence is obtained from a linear fit using the statistical and uncorrelated systematic uncertainties and is found to be $\mathcal{R} = 0.156(6) + 0.0008(6) \sqrt{s}$, where $\sqrt{s}$ is in $\tev$. The observed trend is in agreement with the LHCb measurement of the $f_s/f_u$ dependence upon the $pp$ collision energy~\cite{LHCb-PAPER-2019-020}.
The values for $\mathcal{R}$ will be used in a future work and can be used to obtain $f_{s}/f_{d}$ by correcting $\mathcal{R}$ for the relative $D$ branching fractions, the ratio of $B$ lifetimes, the form factor ratio, the contribution from non-factorisable SU(3)-breaking effects and the contribution from the exchange diagram, as given by Eq.~\eqref{eq:fsfd_abs}.

%% file: summary.tex
\section{Summary}
\label{sec:Summary}
A branching fraction measurement of the $\BdDsPi$ decay is performed using \mbox{$pp$ collision} data taken between 2011 and 2016, leading to 
\begin{equation*}
    \BF(\BdDsPi) = \BFBdDsPiPaper,
\end{equation*}
where the first uncertainty is statistical, the second systematic and the third is due the branching fractions used as normalisation inputs. This is the most precise single measurement of $\BF(\BdDsPi)$ to date, and is in agreement with the current world average~\cite{PDG2020}. Using this branching fraction, the product of $|\Vub|$ and the non-factorisation constant $|\aNF|$ is determined to be
\begin{equation*}
    |\Vub| |\aNF| = \VubaNFExpanded.
\end{equation*}
Comparison with independently measured values of $\Vub$~\cite{PDG2020} indicate that $|\aNF|$ may deviate from unity by around $20\%$, indicating significant non-factorisable corrections.\\
\indent The measurement of the ratio of the $\BdDsPi$ and $\BdDPi$ branching fractions is used to determine the $r_{D\pi}$ parameter, 
\begin{equation*} 
r_{D\pi} = \rDPiExpanded,
\end{equation*}
where the first uncertainty is statistical, the second systematic and the third arises from possible non-factorisable \grpsuthree-breaking effects, estimated to be $20\%$~\cite{DeBruyn:2012jp}. Knowledge of this parameter is essential to interpret the \CP asymmetries in $\decay{\Bd}{\Dmp\pipm}$ decays. \\
\indent Finally, the efficiency-corrected yield ratio of $\BsDsPi$ and $\BdDPi$ decays, $\mathcal{R}$, is used to probe the collision energy dependence of the hadronisation fraction $f_{s}/f_{d}$. 

%% file: acknowledgements.tex
\section*{Acknowledgements}

\noindent We express our gratitude to our colleagues in the CERN
accelerator departments for the excellent performance of the LHC. We
thank the technical and administrative staff at the LHCb
institutes.
We acknowledge support from CERN and from the national agencies:
CAPES, CNPq, FAPERJ and FINEP (Brazil); 
MOST and NSFC (China); 
CNRS/IN2P3 (France); 
BMBF, DFG and MPG (Germany); 
INFN (Italy); 
NWO (Netherlands); 
MNiSW and NCN (Poland); 
MEN/IFA (Romania); 
MSHE (Russia); 
MICINN (Spain); 
SNSF and SER (Switzerland); 
NASU (Ukraine); 
STFC (United Kingdom); 
DOE NP and NSF (USA).
We acknowledge the computing resources that are provided by CERN, IN2P3
(France), KIT and DESY (Germany), INFN (Italy), SURF (Netherlands),
PIC (Spain), GridPP (United Kingdom), RRCKI and Yandex
LLC (Russia), CSCS (Switzerland), IFIN-HH (Romania), CBPF (Brazil),
PL-GRID (Poland) and OSC (USA).
We are indebted to the communities behind the multiple open-source
software packages on which we depend.
Individual groups or members have received support from
AvH Foundation (Germany);
EPLANET, Marie Sk\l{}odowska-Curie Actions and ERC (European Union);
A*MIDEX, ANR, Labex P2IO and OCEVU, and R\'{e}gion Auvergne-Rh\^{o}ne-Alpes (France);
Key Research Program of Frontier Sciences of CAS, CAS PIFI,
Thousand Talents Program, and Sci. \& Tech. Program of Guangzhou (China);
RFBR, RSF and Yandex LLC (Russia);
GVA, XuntaGal and GENCAT (Spain);
the Royal Society
and the Leverhulme Trust (United Kingdom).

%% file: LHCb_Authorship_2020-Oct-22.tex
\centerline
{\large\bf LHCb collaboration}
\begin
{flushleft}
\small
R.~Aaij$^{32}$,
C.~Abell{\'a}n~Beteta$^{50}$,
T.~Ackernley$^{60}$,
B.~Adeva$^{46}$,
M.~Adinolfi$^{54}$,
H.~Afsharnia$^{9}$,
C.A.~Aidala$^{85}$,
S.~Aiola$^{25}$,
Z.~Ajaltouni$^{9}$,
S.~Akar$^{65}$
% ...

{\footnotesize \it

% ...
$^{9}$Universit{\'e} Clermont Auvergne, CNRS/IN2P3, LPC, Clermont-Ferrand, France\\
% ...
$^{25}$INFN Sezione di Milano, Milano, Italy\\
% ...
$^{32}$Nikhef National Institute for Subatomic Physics, Amsterdam, Netherlands\\
% ...
$^{46}$Instituto Galego de F{\'\i}sica de Altas Enerx{\'\i}as (IGFAE), Universidade de Santiago de Compostela, Santiago de Compostela, Spain\\
% ...
$^{50}$Physik-Institut, Universit{\"a}t Z{\"u}rich, Z{\"u}rich, Switzerland\\
% ...
$^{54}$H.H. Wills Physics Laboratory, University of Bristol, Bristol, United Kingdom\\
% ...
$^{60}$Oliver Lodge Laboratory, University of Liverpool, Liverpool, United Kingdom\\
% ...
$^{65}$University of Cincinnati, Cincinnati, OH, United States\\
% ...
$^{68}$Syracuse University, Syracuse, NY, United States\\
% ...
$^{85}$University of Michigan, Ann Arbor, United States, associated to $^{68}$\\
\bigskip
% ...
\medskip
}
\end{flushleft}

%% file: main.bbl
\ifx\mcitethebibliography\mciteundefinedmacro
\PackageError{LHCb.bst}{mciteplus.sty has not been loaded}
{This bibstyle requires the use of the mciteplus package.}\fi
\providecommand{\href}[2]{#2}
\begin{mcitethebibliography}{10}
\mciteSetBstSublistMode{n}
\mciteSetBstMaxWidthForm{subitem}{\alph{mcitesubitemcount})}
\mciteSetBstSublistLabelBeginEnd{\mcitemaxwidthsubitemform\space}
{\relax}{\relax}

\bibitem{CKMfitter2015}
CKMfitter group, J.~Charles {\em et~al.},
  \ifthenelse{\boolean{articletitles}}{\emph{{Current status of the standard
  model CKM fit and constraints on \hbox{$\Delta F=2$} new physics}},
  }{}\href{https://doi.org/10.1103/PhysRevD.91.073007}{Phys.\ Rev.\
  \textbf{D91} (2015) 073007},
  \href{http://arxiv.org/abs/1501.05013}{{\normalfont\ttfamily
  arXiv:1501.05013}}, {updated results and plots available at
  \href{http://ckmfitter.in2p3.fr/}{{\texttt{http://ckmfitter.in2p3.fr/}}}}\relax
\mciteBstWouldAddEndPuncttrue
\mciteSetBstMidEndSepPunct{\mcitedefaultmidpunct}
{\mcitedefaultendpunct}{\mcitedefaultseppunct}\relax
\EndOfBibitem
\bibitem{UTfit-UT}
UTfit collaboration, M.~Bona {\em et~al.},
  \ifthenelse{\boolean{articletitles}}{\emph{{The unitarity triangle fit in the
  standard model and hadronic parameters from lattice QCD: A reappraisal after
  the measurements of $\Delta m_{s}$ and $BR(B\to\tau\nu_{\tau})$}},
  }{}\href{https://doi.org/10.1088/1126-6708/2006/10/081}{JHEP \textbf{10}
  (2006) 081}, \href{http://arxiv.org/abs/hep-ph/0606167}{{\normalfont\ttfamily
  arXiv:hep-ph/0606167}}, {updated results and plots available at
  \href{http://www.utfit.org/}{{\texttt{http://www.utfit.org/}}}}\relax
\mciteBstWouldAddEndPuncttrue
\mciteSetBstMidEndSepPunct{\mcitedefaultmidpunct}
{\mcitedefaultendpunct}{\mcitedefaultseppunct}\relax
\EndOfBibitem
\bibitem{formfactor}
P.~Ball and R.~Zwicky, \ifthenelse{\boolean{articletitles}}{\emph{{New results
  on $B \to \pi, K, \eta$ decay form factors from light-cone sum rules}},
  }{}\href{https://doi.org/10.1103/PhysRevD.71.014015}{Phys.\ Rev.\
  \textbf{D71} (2005) 014015},
  \href{http://arxiv.org/abs/hep-ph/0406232}{{\normalfont\ttfamily
  arXiv:hep-ph/0406232}}\relax
\mciteBstWouldAddEndPuncttrue
\mciteSetBstMidEndSepPunct{\mcitedefaultmidpunct}
{\mcitedefaultendpunct}{\mcitedefaultseppunct}\relax
\EndOfBibitem
\bibitem{Ball:2006jz}
P.~Ball, \ifthenelse{\boolean{articletitles}}{\emph{{$|\Vub|$ from UTangles and
  $\Bd \to \pim \ellp \nu$}},
  }{}\href{https://doi.org/10.1016/j.physletb.2006.11.034}{Phys.\ Lett.\
  \textbf{B644} (2007) 38},
  \href{http://arxiv.org/abs/hep-ph/0611108}{{\normalfont\ttfamily
  arXiv:hep-ph/0611108}}\relax
\mciteBstWouldAddEndPuncttrue
\mciteSetBstMidEndSepPunct{\mcitedefaultmidpunct}
{\mcitedefaultendpunct}{\mcitedefaultseppunct}\relax
\EndOfBibitem
\bibitem{decayconstant}
A.~Bazavov {\em et~al.}, \ifthenelse{\boolean{articletitles}}{\emph{{$B$- and
  $D$-meson leptonic decay constants from four-flavor lattice QCD}},
  }{}\href{https://doi.org/10.1103/PhysRevD.98.074512}{Phys.\ Rev.\
  \textbf{D98} (2018) 074512},
  \href{http://arxiv.org/abs/1712.09262}{{\normalfont\ttfamily
  arXiv:1712.09262}}\relax
\mciteBstWouldAddEndPuncttrue
\mciteSetBstMidEndSepPunct{\mcitedefaultmidpunct}
{\mcitedefaultendpunct}{\mcitedefaultseppunct}\relax
\EndOfBibitem
\bibitem{decayconstant2}
N.~Carrasco {\em et~al.}, \ifthenelse{\boolean{articletitles}}{\emph{{Leptonic
  decay constants $f_{K},f_{D},$ and $f_{{D}_{s}}$ with $N_{f} = 2+1+1$
  twisted-mass lattice QCD}},
  }{}\href{https://doi.org/10.1103/PhysRevD.91.054507}{Phys.\ Rev.\
  \textbf{D91} (2015) 054507},
  \href{http://arxiv.org/abs/1411.7908}{{\normalfont\ttfamily
  arXiv:1411.7908}}\relax
\mciteBstWouldAddEndPuncttrue
\mciteSetBstMidEndSepPunct{\mcitedefaultmidpunct}
{\mcitedefaultendpunct}{\mcitedefaultseppunct}\relax
\EndOfBibitem
\bibitem{Beneke:2000ry}
M.~Beneke, G.~Buchalla, M.~Neubert, and C.~T. Sachrajda,
  \ifthenelse{\boolean{articletitles}}{\emph{{QCD factorization for exclusive,
  nonleptonic B meson decays: General arguments and the case of heavy light
  final states}},
  }{}\href{https://doi.org/10.1016/S0550-3213(00)00559-9}{Nucl.\ Phys.\
  \textbf{B591} (2000) 313},
  \href{http://arxiv.org/abs/hep-ph/0006124}{{\normalfont\ttfamily
  arXiv:hep-ph/0006124}}\relax
\mciteBstWouldAddEndPuncttrue
\mciteSetBstMidEndSepPunct{\mcitedefaultmidpunct}
{\mcitedefaultendpunct}{\mcitedefaultseppunct}\relax
\EndOfBibitem
\bibitem{Aubert:2005yf}
BaBar collaboration, B.~Aubert {\em et~al.},
  \ifthenelse{\boolean{articletitles}}{\emph{{Measurement of time-dependent
  CP-violating asymmetries and constraints on $\sin(2\beta+\gamma)$ with
  partial reconstruction of $B \to D^{*\mp} \pi^\pm$ decays}},
  }{}\href{https://doi.org/10.1103/PhysRevD.71.112003}{Phys.\ Rev.\
  \textbf{D71} (2005) 112003},
  \href{http://arxiv.org/abs/hep-ex/0504035}{{\normalfont\ttfamily
  arXiv:hep-ex/0504035}}\relax
\mciteBstWouldAddEndPuncttrue
\mciteSetBstMidEndSepPunct{\mcitedefaultmidpunct}
{\mcitedefaultendpunct}{\mcitedefaultseppunct}\relax
\EndOfBibitem
\bibitem{Aubert:2006tw}
BaBar collaboration, B.~Aubert {\em et~al.},
  \ifthenelse{\boolean{articletitles}}{\emph{{Measurement of time-dependent CP
  asymmetries in $B^0 \to D^{(*)\pm}$ $\pi^\mp$ and $B^0 \to D^\pm \rho^\mp$
  decays}}, }{}\href{https://doi.org/10.1103/PhysRevD.73.111101}{Phys.\ Rev.\
  \textbf{D73} (2006) 111101},
  \href{http://arxiv.org/abs/hep-ex/0602049}{{\normalfont\ttfamily
  arXiv:hep-ex/0602049}}\relax
\mciteBstWouldAddEndPuncttrue
\mciteSetBstMidEndSepPunct{\mcitedefaultmidpunct}
{\mcitedefaultendpunct}{\mcitedefaultseppunct}\relax
\EndOfBibitem
\bibitem{Ronga:2006hv}
Belle collaboration, F.~J. Ronga {\em et~al.},
  \ifthenelse{\boolean{articletitles}}{\emph{{Measurements of CP violation in
  $\Bd\rightarrow D^{*-}\pip$ and $\Bd \rightarrow \Dm\pip$ decays}},
  }{}\href{https://doi.org/10.1103/PhysRevD.73.092003}{Phys.\ Rev.\
  \textbf{D73} (2006) 092003},
  \href{http://arxiv.org/abs/hep-ex/0604013}{{\normalfont\ttfamily
  arXiv:hep-ex/0604013}}\relax
\mciteBstWouldAddEndPuncttrue
\mciteSetBstMidEndSepPunct{\mcitedefaultmidpunct}
{\mcitedefaultendpunct}{\mcitedefaultseppunct}\relax
\EndOfBibitem
\bibitem{Bahinipati:2011yq}
Belle collaboration, S.~Bahinipati {\em et~al.},
  \ifthenelse{\boolean{articletitles}}{\emph{{Measurements of time-dependent CP
  asymmetries in $B \to D^{*\mp} \pi^{\pm}$ decays using a partial
  reconstruction technique}},
  }{}\href{https://doi.org/10.1103/PhysRevD.84.021101}{Phys.\ Rev.\
  \textbf{D84} (2011) 021101},
  \href{http://arxiv.org/abs/1102.0888}{{\normalfont\ttfamily
  arXiv:1102.0888}}\relax
\mciteBstWouldAddEndPuncttrue
\mciteSetBstMidEndSepPunct{\mcitedefaultmidpunct}
{\mcitedefaultendpunct}{\mcitedefaultseppunct}\relax
\EndOfBibitem
\bibitem{DeBruyn:2012jp}
K.~De~Bruyn {\em et~al.}, \ifthenelse{\boolean{articletitles}}{\emph{{Exploring
  $B_s \to D_s^{(*)\pm} K^\mp$ decays in the presence of a sizable width
  difference $\Delta\Gamma_s$}},
  }{}\href{https://doi.org/10.1016/j.nuclphysb.2012.11.012}{Nucl.\ Phys.\
  \textbf{B868} (2013) 351},
  \href{http://arxiv.org/abs/1208.6463}{{\normalfont\ttfamily
  arXiv:1208.6463}}\relax
\mciteBstWouldAddEndPuncttrue
\mciteSetBstMidEndSepPunct{\mcitedefaultmidpunct}
{\mcitedefaultendpunct}{\mcitedefaultseppunct}\relax
\EndOfBibitem
\bibitem{LHCb-PAPER-2018-009}
LHCb collaboration, R.~Aaij {\em et~al.},
  \ifthenelse{\boolean{articletitles}}{\emph{{Measurement of \CP violation in
  \mbox{\decay{\Bz}{D^\pm \pimp}} decays}},
  }{}\href{https://doi.org/10.1007/JHEP06(2018)084}{JHEP \textbf{06} (2018)
  084}, \href{http://arxiv.org/abs/1805.03448}{{\normalfont\ttfamily
  arXiv:1805.03448}}\relax
\mciteBstWouldAddEndPuncttrue
\mciteSetBstMidEndSepPunct{\mcitedefaultmidpunct}
{\mcitedefaultendpunct}{\mcitedefaultseppunct}\relax
\EndOfBibitem
\bibitem{Aubert:2008zi}
BaBar collaboration, B.~Aubert {\em et~al.},
  \ifthenelse{\boolean{articletitles}}{\emph{{Measurement of the branching
  fractions of the rare decays $B^0 \to D_{s}^{(*)+} \pim$, $B^0 \to
  D_{s}^{(*)+} \rho^{-}$, and $B^0 \to D_{s}^{(*)-} K^{(*)+}$}},
  }{}\href{https://doi.org/10.1103/PhysRevD.78.032005}{Phys.\ Rev.\
  \textbf{D78} (2008) 032005},
  \href{http://arxiv.org/abs/0803.4296}{{\normalfont\ttfamily
  arXiv:0803.4296}}\relax
\mciteBstWouldAddEndPuncttrue
\mciteSetBstMidEndSepPunct{\mcitedefaultmidpunct}
{\mcitedefaultendpunct}{\mcitedefaultseppunct}\relax
\EndOfBibitem
\bibitem{Das:2010be}
Belle collaboration, A.~Das {\em et~al.},
  \ifthenelse{\boolean{articletitles}}{\emph{{Measurements of branching
  fractions for \mbox{$\BdDsPi$} and $\bar{B}^0 \rightarrow D_s^+K^-$}},
  }{}\href{https://doi.org/10.1103/PhysRevD.82.051103}{Phys.\ Rev.\
  \textbf{D82} (2010) 051103},
  \href{http://arxiv.org/abs/1007.4619}{{\normalfont\ttfamily
  arXiv:1007.4619}}\relax
\mciteBstWouldAddEndPuncttrue
\mciteSetBstMidEndSepPunct{\mcitedefaultmidpunct}
{\mcitedefaultendpunct}{\mcitedefaultseppunct}\relax
\EndOfBibitem
\bibitem{LHCb-PAPER-2019-020}
LHCb collaboration, R.~Aaij {\em et~al.},
  \ifthenelse{\boolean{articletitles}}{\emph{{Measurement of $f_s / f_u$
  variation with proton-proton collision energy and \B-meson kinematics}},
  }{}\href{https://doi.org/10.1103/PhysRevLett.124.122002}{Phys.\ Rev.\ Lett.\
  \textbf{124} (2020) 122002},
  \href{http://arxiv.org/abs/1910.09934}{{\normalfont\ttfamily
  arXiv:1910.09934}}\relax
\mciteBstWouldAddEndPuncttrue
\mciteSetBstMidEndSepPunct{\mcitedefaultmidpunct}
{\mcitedefaultendpunct}{\mcitedefaultseppunct}\relax
\EndOfBibitem
\bibitem{LHCb-PAPER-2017-001}
LHCb collaboration, R.~Aaij {\em et~al.},
  \ifthenelse{\boolean{articletitles}}{\emph{{Measurement of the
  \mbox{\decay{\Bs}{\mumu}} branching fraction and effective lifetime and
  search for \mbox{\decay{\Bz}{\mumu}} decays}},
  }{}\href{https://doi.org/10.1103/PhysRevLett.118.191801}{Phys.\ Rev.\ Lett.\
  \textbf{118} (2017) 191801},
  \href{http://arxiv.org/abs/1703.05747}{{\normalfont\ttfamily
  arXiv:1703.05747}}\relax
\mciteBstWouldAddEndPuncttrue
\mciteSetBstMidEndSepPunct{\mcitedefaultmidpunct}
{\mcitedefaultendpunct}{\mcitedefaultseppunct}\relax
\EndOfBibitem
\bibitem{Fleischer:2010ca}
R.~Fleischer, N.~Serra, and N.~Tuning,
  \ifthenelse{\boolean{articletitles}}{\emph{{Tests of factorization and SU(3)
  relations in B decays into heavy-light final states}},
  }{}\href{https://doi.org/10.1103/PhysRevD.83.014017}{Phys.\ Rev.\
  \textbf{D83} (2011) 014017},
  \href{http://arxiv.org/abs/1012.2784}{{\normalfont\ttfamily
  arXiv:1012.2784}}\relax
\mciteBstWouldAddEndPuncttrue
\mciteSetBstMidEndSepPunct{\mcitedefaultmidpunct}
{\mcitedefaultendpunct}{\mcitedefaultseppunct}\relax
\EndOfBibitem
\bibitem{LHCb-DP-2008-001}
LHCb collaboration, A.~A. Alves~Jr.\ {\em et~al.},
  \ifthenelse{\boolean{articletitles}}{\emph{{The \lhcb detector at the LHC}},
  }{}\href{https://doi.org/10.1088/1748-0221/3/08/S08005}{JINST \textbf{3}
  (2008) S08005}\relax
\mciteBstWouldAddEndPuncttrue
\mciteSetBstMidEndSepPunct{\mcitedefaultmidpunct}
{\mcitedefaultendpunct}{\mcitedefaultseppunct}\relax
\EndOfBibitem
\bibitem{LHCb-DP-2014-002}
LHCb collaboration, R.~Aaij {\em et~al.},
  \ifthenelse{\boolean{articletitles}}{\emph{{LHCb detector performance}},
  }{}\href{https://doi.org/10.1142/S0217751X15300227}{Int.\ J.\ Mod.\ Phys.\
  \textbf{A30} (2015) 1530022},
  \href{http://arxiv.org/abs/1412.6352}{{\normalfont\ttfamily
  arXiv:1412.6352}}\relax
\mciteBstWouldAddEndPuncttrue
\mciteSetBstMidEndSepPunct{\mcitedefaultmidpunct}
{\mcitedefaultendpunct}{\mcitedefaultseppunct}\relax
\EndOfBibitem
\bibitem{LHCb-DP-2014-001}
R.~Aaij {\em et~al.}, \ifthenelse{\boolean{articletitles}}{\emph{{Performance
  of the LHCb Vertex Locator}},
  }{}\href{https://doi.org/10.1088/1748-0221/9/09/P09007}{JINST \textbf{9}
  (2014) P09007}, \href{http://arxiv.org/abs/1405.7808}{{\normalfont\ttfamily
  arXiv:1405.7808}}\relax
\mciteBstWouldAddEndPuncttrue
\mciteSetBstMidEndSepPunct{\mcitedefaultmidpunct}
{\mcitedefaultendpunct}{\mcitedefaultseppunct}\relax
\EndOfBibitem
\bibitem{LHCb-DP-2013-003}
R.~Arink {\em et~al.}, \ifthenelse{\boolean{articletitles}}{\emph{{Performance
  of the LHCb Outer Tracker}},
  }{}\href{https://doi.org/10.1088/1748-0221/9/01/P01002}{JINST \textbf{9}
  (2014) P01002}, \href{http://arxiv.org/abs/1311.3893}{{\normalfont\ttfamily
  arXiv:1311.3893}}\relax
\mciteBstWouldAddEndPuncttrue
\mciteSetBstMidEndSepPunct{\mcitedefaultmidpunct}
{\mcitedefaultendpunct}{\mcitedefaultseppunct}\relax
\EndOfBibitem
\bibitem{LHCb-DP-2017-001}
P.~d'Argent {\em et~al.}, \ifthenelse{\boolean{articletitles}}{\emph{{Improved
  performance of the LHCb Outer Tracker in LHC Run 2}},
  }{}\href{https://doi.org/10.1088/1748-0221/12/11/P11016}{JINST \textbf{12}
  (2017) P11016}, \href{http://arxiv.org/abs/1708.00819}{{\normalfont\ttfamily
  arXiv:1708.00819}}\relax
\mciteBstWouldAddEndPuncttrue
\mciteSetBstMidEndSepPunct{\mcitedefaultmidpunct}
{\mcitedefaultendpunct}{\mcitedefaultseppunct}\relax
\EndOfBibitem
\bibitem{LHCb-DP-2012-003}
M.~Adinolfi {\em et~al.},
  \ifthenelse{\boolean{articletitles}}{\emph{{Performance of the \lhcb RICH
  detector at the LHC}},
  }{}\href{https://doi.org/10.1140/epjc/s10052-013-2431-9}{Eur.\ Phys.\ J.\
  \textbf{C73} (2013) 2431},
  \href{http://arxiv.org/abs/1211.6759}{{\normalfont\ttfamily
  arXiv:1211.6759}}\relax
\mciteBstWouldAddEndPuncttrue
\mciteSetBstMidEndSepPunct{\mcitedefaultmidpunct}
{\mcitedefaultendpunct}{\mcitedefaultseppunct}\relax
\EndOfBibitem
\bibitem{LHCb-DP-2012-002}
A.~A. Alves~Jr.\ {\em et~al.},
  \ifthenelse{\boolean{articletitles}}{\emph{{Performance of the LHCb muon
  system}}, }{}\href{https://doi.org/10.1088/1748-0221/8/02/P02022}{JINST
  \textbf{8} (2013) P02022},
  \href{http://arxiv.org/abs/1211.1346}{{\normalfont\ttfamily
  arXiv:1211.1346}}\relax
\mciteBstWouldAddEndPuncttrue
\mciteSetBstMidEndSepPunct{\mcitedefaultmidpunct}
{\mcitedefaultendpunct}{\mcitedefaultseppunct}\relax
\EndOfBibitem
\bibitem{LHCb-DP-2012-004}
R.~Aaij {\em et~al.}, \ifthenelse{\boolean{articletitles}}{\emph{{The \lhcb
  trigger and its performance in 2011}},
  }{}\href{https://doi.org/10.1088/1748-0221/8/04/P04022}{JINST \textbf{8}
  (2013) P04022}, \href{http://arxiv.org/abs/1211.3055}{{\normalfont\ttfamily
  arXiv:1211.3055}}\relax
\mciteBstWouldAddEndPuncttrue
\mciteSetBstMidEndSepPunct{\mcitedefaultmidpunct}
{\mcitedefaultendpunct}{\mcitedefaultseppunct}\relax
\EndOfBibitem
\bibitem{Sjostrand:2007gs}
T.~Sj\"{o}strand, S.~Mrenna, and P.~Skands,
  \ifthenelse{\boolean{articletitles}}{\emph{{A brief introduction to PYTHIA
  8.1}}, }{}\href{https://doi.org/10.1016/j.cpc.2008.01.036}{Comput.\ Phys.\
  Commun.\  \textbf{178} (2008) 852},
  \href{http://arxiv.org/abs/0710.3820}{{\normalfont\ttfamily
  arXiv:0710.3820}}\relax
\mciteBstWouldAddEndPuncttrue
\mciteSetBstMidEndSepPunct{\mcitedefaultmidpunct}
{\mcitedefaultendpunct}{\mcitedefaultseppunct}\relax
\EndOfBibitem
\bibitem{LHCb-PROC-2010-056}
I.~Belyaev {\em et~al.}, \ifthenelse{\boolean{articletitles}}{\emph{{Handling
  of the generation of primary events in Gauss, the LHCb simulation
  framework}}, }{}\href{https://doi.org/10.1088/1742-6596/331/3/032047}{J.\
  Phys.\ Conf.\ Ser.\  \textbf{331} (2011) 032047}\relax
\mciteBstWouldAddEndPuncttrue
\mciteSetBstMidEndSepPunct{\mcitedefaultmidpunct}
{\mcitedefaultendpunct}{\mcitedefaultseppunct}\relax
\EndOfBibitem
\bibitem{Lange:2001uf}
D.~J. Lange, \ifthenelse{\boolean{articletitles}}{\emph{{The EvtGen particle
  decay simulation package}},
  }{}\href{https://doi.org/10.1016/S0168-9002(01)00089-4}{Nucl.\ Instrum.\
  Meth.\  \textbf{A462} (2001) 152}\relax
\mciteBstWouldAddEndPuncttrue
\mciteSetBstMidEndSepPunct{\mcitedefaultmidpunct}
{\mcitedefaultendpunct}{\mcitedefaultseppunct}\relax
\EndOfBibitem
\bibitem{Golonka:2005pn}
P.~Golonka and Z.~Was, \ifthenelse{\boolean{articletitles}}{\emph{{PHOTOS Monte
  Carlo: A precision tool for QED corrections in $Z$ and $W$ decays}},
  }{}\href{https://doi.org/10.1140/epjc/s2005-02396-4}{Eur.\ Phys.\ J.\
  \textbf{C45} (2006) 97},
  \href{http://arxiv.org/abs/hep-ph/0506026}{{\normalfont\ttfamily
  arXiv:hep-ph/0506026}}\relax
\mciteBstWouldAddEndPuncttrue
\mciteSetBstMidEndSepPunct{\mcitedefaultmidpunct}
{\mcitedefaultendpunct}{\mcitedefaultseppunct}\relax
\EndOfBibitem
\bibitem{Allison:2006ve}
Geant4 collaboration, J.~Allison {\em et~al.},
  \ifthenelse{\boolean{articletitles}}{\emph{{Geant4 developments and
  applications}}, }{}\href{https://doi.org/10.1109/TNS.2006.869826}{IEEE
  Trans.\ Nucl.\ Sci.\  \textbf{53} (2006) 270}\relax
\mciteBstWouldAddEndPuncttrue
\mciteSetBstMidEndSepPunct{\mcitedefaultmidpunct}
{\mcitedefaultendpunct}{\mcitedefaultseppunct}\relax
\EndOfBibitem
\bibitem{Agostinelli:2002hh}
Geant4 collaboration, S.~Agostinelli {\em et~al.},
  \ifthenelse{\boolean{articletitles}}{\emph{{Geant4: A simulation toolkit}},
  }{}\href{https://doi.org/10.1016/S0168-9002(03)01368-8}{Nucl.\ Instrum.\
  Meth.\  \textbf{A506} (2003) 250}\relax
\mciteBstWouldAddEndPuncttrue
\mciteSetBstMidEndSepPunct{\mcitedefaultmidpunct}
{\mcitedefaultendpunct}{\mcitedefaultseppunct}\relax
\EndOfBibitem
\bibitem{LHCb-PROC-2011-006}
M.~Clemencic {\em et~al.}, \ifthenelse{\boolean{articletitles}}{\emph{{The
  \lhcb simulation application, Gauss: Design, evolution and experience}},
  }{}\href{https://doi.org/10.1088/1742-6596/331/3/032023}{J.\ Phys.\ Conf.\
  Ser.\  \textbf{331} (2011) 032023}\relax
\mciteBstWouldAddEndPuncttrue
\mciteSetBstMidEndSepPunct{\mcitedefaultmidpunct}
{\mcitedefaultendpunct}{\mcitedefaultseppunct}\relax
\EndOfBibitem
\bibitem{BBDT}
V.~V. Gligorov and M.~Williams,
  \ifthenelse{\boolean{articletitles}}{\emph{{Efficient, reliable and fast
  high-level triggering using a bonsai boosted decision tree}},
  }{}\href{https://doi.org/10.1088/1748-0221/8/02/P02013}{JINST \textbf{8}
  (2013) P02013}, \href{http://arxiv.org/abs/1210.6861}{{\normalfont\ttfamily
  arXiv:1210.6861}}\relax
\mciteBstWouldAddEndPuncttrue
\mciteSetBstMidEndSepPunct{\mcitedefaultmidpunct}
{\mcitedefaultendpunct}{\mcitedefaultseppunct}\relax
\EndOfBibitem
\bibitem{Breiman}
L.~Breiman, J.~H. Friedman, R.~A. Olshen, and C.~J. Stone, {\em Classification
  and regression trees}, Wadsworth international group, Belmont, California,
  USA, 1984\relax
\mciteBstWouldAddEndPuncttrue
\mciteSetBstMidEndSepPunct{\mcitedefaultmidpunct}
{\mcitedefaultendpunct}{\mcitedefaultseppunct}\relax
\EndOfBibitem
\bibitem{Roe:2004na}
B.~P. Roe {\em et~al.}, \ifthenelse{\boolean{articletitles}}{\emph{{Boosted
  decision trees, an alternative to artificial neural networks}},
  }{}\href{https://doi.org/10.1016/j.nima.2004.12.018}{Nucl.\ Instrum.\ Meth.\
  \textbf{A543} (2005) 577},
  \href{http://arxiv.org/abs/physics/0408124}{{\normalfont\ttfamily
  arXiv:physics/0408124}}\relax
\mciteBstWouldAddEndPuncttrue
\mciteSetBstMidEndSepPunct{\mcitedefaultmidpunct}
{\mcitedefaultendpunct}{\mcitedefaultseppunct}\relax
\EndOfBibitem
\bibitem{Eitschberger:2018ofp}
U.~P. Eitschberger, {\em {Flavour-tagged measurement of CP observables in
  $B^0_s \to D_s^\mp K^\pm$ decays with the LHCb experiment}}, PhD thesis,
  Tech. U., Dortmund (main), 2018,
  doi:~\href{https://doi.org/10.17877/DE290R-18881}{10.17877/DE290R-18881}\relax
\mciteBstWouldAddEndPuncttrue
\mciteSetBstMidEndSepPunct{\mcitedefaultmidpunct}
{\mcitedefaultendpunct}{\mcitedefaultseppunct}\relax
\EndOfBibitem
\bibitem{PDG2020}
Particle Data Group, P.~A. Zyla {\em et~al.},
  \ifthenelse{\boolean{articletitles}}{\emph{{\href{http://pdg.lbl.gov/}{Review
  of particle physics}}}, }{}\href{https://doi.org/10.1093/ptep/ptaa104}{Prog.\
  Theor.\ Exp.\ Phys.\  \textbf{2020} (2020) 083C01}\relax
\mciteBstWouldAddEndPuncttrue
\mciteSetBstMidEndSepPunct{\mcitedefaultmidpunct}
{\mcitedefaultendpunct}{\mcitedefaultseppunct}\relax
\EndOfBibitem
\bibitem{LHCb-PAPER-2017-047}
LHCb collaboration, R.~Aaij {\em et~al.},
  \ifthenelse{\boolean{articletitles}}{\emph{{Measurement of \CP asymmetry in
  \mbox{\decay{\Bs}{D_s^\mp \Kpm}} decays}},
  }{}\href{https://doi.org/10.1007/JHEP03(2018)059}{JHEP \textbf{03} (2018)
  059}, \href{http://arxiv.org/abs/1712.07428}{{\normalfont\ttfamily
  arXiv:1712.07428}}\relax
\mciteBstWouldAddEndPuncttrue
\mciteSetBstMidEndSepPunct{\mcitedefaultmidpunct}
{\mcitedefaultendpunct}{\mcitedefaultseppunct}\relax
\EndOfBibitem
\bibitem{Santos:2013gra}
D.~Mart{\'\i}nez~Santos and F.~Dupertuis,
  \ifthenelse{\boolean{articletitles}}{\emph{{Mass distributions marginalized
  over per-event errors}},
  }{}\href{https://doi.org/10.1016/j.nima.2014.06.081}{Nucl.\ Instrum.\ Meth.\
  \textbf{A764} (2014) 150},
  \href{http://arxiv.org/abs/1312.5000}{{\normalfont\ttfamily
  arXiv:1312.5000}}\relax
\mciteBstWouldAddEndPuncttrue
\mciteSetBstMidEndSepPunct{\mcitedefaultmidpunct}
{\mcitedefaultendpunct}{\mcitedefaultseppunct}\relax
\EndOfBibitem
\bibitem{johnson1949systems}
N.~L. Johnson, \ifthenelse{\boolean{articletitles}}{\emph{Systems of frequency
  curves generated by methods of translation}, }{}Biometrika \textbf{36} (1949)
  149\relax
\mciteBstWouldAddEndPuncttrue
\mciteSetBstMidEndSepPunct{\mcitedefaultmidpunct}
{\mcitedefaultendpunct}{\mcitedefaultseppunct}\relax
\EndOfBibitem
\bibitem{LHCb-PAPER-2017-021}
LHCb collaboration, R.~Aaij {\em et~al.},
  \ifthenelse{\boolean{articletitles}}{\emph{{Measurement of \CP observables in
  \mbox{\decay{\Bpm}{D^{(\ast)}\Kpm}} and \mbox{\decay{\Bpm}{D^{(\ast)}\pipm}}
  decays}}, }{}\href{https://doi.org/10.1016/j.physletb.2017.11.070}{Phys.\
  Lett.\  \textbf{B777} (2018) 16},
  \href{http://arxiv.org/abs/1708.06370}{{\normalfont\ttfamily
  arXiv:1708.06370}}\relax
\mciteBstWouldAddEndPuncttrue
\mciteSetBstMidEndSepPunct{\mcitedefaultmidpunct}
{\mcitedefaultendpunct}{\mcitedefaultseppunct}\relax
\EndOfBibitem
\bibitem{Cranmer:2000du}
K.~S. Cranmer, \ifthenelse{\boolean{articletitles}}{\emph{{Kernel estimation in
  high-energy physics}},
  }{}\href{https://doi.org/10.1016/S0010-4655(00)00243-5}{Comput.\ Phys.\
  Commun.\  \textbf{136} (2001) 198},
  \href{http://arxiv.org/abs/hep-ex/0011057}{{\normalfont\ttfamily
  arXiv:hep-ex/0011057}}\relax
\mciteBstWouldAddEndPuncttrue
\mciteSetBstMidEndSepPunct{\mcitedefaultmidpunct}
{\mcitedefaultendpunct}{\mcitedefaultseppunct}\relax
\EndOfBibitem
\bibitem{LHCb-PAPER-2018-036}
LHCb collaboration, R.~Aaij {\em et~al.},
  \ifthenelse{\boolean{articletitles}}{\emph{{Measurement of the branching
  fraction and \CP asymmetry in \mbox{\decay{\Bp}{\jpsi\rhop}} decays}},
  }{}\href{https://doi.org/10.1140/epjc/s10052-019-6698-3}{Eur.\ Phys.\ J.\
  \textbf{C79} (2019) 537},
  \href{http://arxiv.org/abs/1812.07041}{{\normalfont\ttfamily
  arXiv:1812.07041}}\relax
\mciteBstWouldAddEndPuncttrue
\mciteSetBstMidEndSepPunct{\mcitedefaultmidpunct}
{\mcitedefaultendpunct}{\mcitedefaultseppunct}\relax
\EndOfBibitem
\bibitem{Pivk:2004ty}
M.~Pivk and F.~R. Le~Diberder,
  \ifthenelse{\boolean{articletitles}}{\emph{{sPlot: A statistical tool to
  unfold data distributions}},
  }{}\href{https://doi.org/10.1016/j.nima.2005.08.106}{Nucl.\ Instrum.\ Meth.\
  \textbf{A555} (2005) 356},
  \href{http://arxiv.org/abs/physics/0402083}{{\normalfont\ttfamily
  arXiv:physics/0402083}}\relax
\mciteBstWouldAddEndPuncttrue
\mciteSetBstMidEndSepPunct{\mcitedefaultmidpunct}
{\mcitedefaultendpunct}{\mcitedefaultseppunct}\relax
\EndOfBibitem
\bibitem{LHCb-PAPER-2012-037}
LHCb collaboration, R.~Aaij {\em et~al.},
  \ifthenelse{\boolean{articletitles}}{\emph{{Measurement of the fragmentation
  fraction ratio $f_s/f_d$ and its dependence on \B meson kinematics}},
  }{}\href{https://doi.org/10.1007/JHEP04(2013)001}{JHEP \textbf{04} (2013)
  001}, \href{http://arxiv.org/abs/1301.5286}{{\normalfont\ttfamily
  arXiv:1301.5286}}\relax
\mciteBstWouldAddEndPuncttrue
\mciteSetBstMidEndSepPunct{\mcitedefaultmidpunct}
{\mcitedefaultendpunct}{\mcitedefaultseppunct}\relax
\EndOfBibitem
\end{mcitethebibliography}
